\documentclass[aps,pra,showpacs,amsmath,amssymb,amsfonts,twocolumn,superscriptaddress,floatfix,footinbib]{revtex4}
\usepackage{epsfig}
\usepackage{graphicx}

\newcommand{\bgar}{\begin{eqnarray}}
\newcommand{\enar}{\end{eqnarray}}

\newcommand{\be}{\begin{equation}}
\newcommand{\ee}{\end{equation}}

\newcommand{\Rmat}[1]{\mathbb{#1}}

\begin{document}

\title{\bf 
Dynamical response of the ``GGG'' rotor to test the Equivalence Principle: theory, simulation and experiment. Part I: the normal modes}

\author{G. L. Comandi} 
\affiliation{INFN-Istituto Nazionale di Fisica Nucleare, Sezione di Pisa, Largo B. Pontecorvo 3,  I-56127 Pisa, Italy}
\affiliation{Dept. of Physics, University of Bologna, Bologna, Italy}

\author{M. L. Chiofalo}
\affiliation{Scuola Normale Superiore, Piazza dei Cavalieri 7, I-56100 Pisa, Italy}
\affiliation{INFN-Istituto Nazionale di Fisica Nucleare, Sezione di Pisa, Largo B. Pontecorvo 3,  I-56127 Pisa, Italy}
\affiliation{INFM-Istituto Nazionale per la Fisica della Materia, Italy}

\author{R. Toncelli} 
\affiliation{INFN-Istituto Nazionale di Fisica Nucleare, Sezione di Pisa, Largo B. Pontecorvo 3,  I-56127 Pisa, Italy}

\author{ D.  Bramanti}
\affiliation{INFN-Istituto Nazionale di Fisica Nucleare, Sezione di Pisa, Largo B. Pontecorvo 3,  I-56127 Pisa, Italy}

\author{E. Polacco}
\affiliation{
Dept. of Physics ``E. Fermi'', University of Pisa, Largo B. Pontecorvo 3,
I-56127 Pisa, Italy}
\affiliation{INFN-Istituto Nazionale di Fisica Nucleare, Sezione di Pisa, Largo B. Pontecorvo 3,  I-56127 Pisa, Italy}

\author{A. M. Nobili}
\affiliation{
Dept. of Physics ``E. Fermi'', University of Pisa, Largo B. Pontecorvo 3,
I-56127 Pisa, Italy}
\affiliation{INFN-Istituto Nazionale di Fisica Nucleare, Sezione di Pisa, Largo B. Pontecorvo 3,  I-56127 Pisa, Italy}

\begin{abstract}

Recent theoretical work suggests that violation of the Equivalence Principle  might be revealed in a measurement of  the fractional differential acceleration $\eta$ between  two test bodies $\--$of different composition, falling in the gravitational field of a source mass$\--$ if the measurement is made to the level of $\eta\simeq 10^{-13}$ or better. This being within the reach of ground based experiments, gives them a new impetus. However, while slowly rotating torsion balances in ground laboratories are close to reaching this level, only an experiment performed in low orbit around the Earth is likely to provide a much better accuracy.  

We report on the progress made with the ``Galileo Galilei on the Ground'' (GGG) experiment, which aims to compete with torsion balances  using an instrument  design also capable of being  converted into a much higher sensitivity space test.

In the present and following paper (Part I and Part II), we demonstrate that the dynamical response of the GGG differential accelerometer set into supercritical rotation $\--$in particular its normal modes (Part I) and rejection of common mode effects (Part II)$\--$ can be predicted by means of a simple but effective model that embodies all the relevant physics. Analytical solutions are obtained under special limits, which provide the theoretical understanding. A simulation environment is set up, obtaining quantitative agreement with the available experimental data on the frequencies of the normal modes, and on the whirling behavior. This is a needed and reliable tool for controlling  and separating perturbative effects from the expected signal, as well as for planning the optimization of the apparatus.

\end{abstract}

\pacs{04.80.Cc, 07.10.-h, 06.30.Bp, 07.87.+v}
\maketitle

\section{Introduction}\label{SecInt} 

Experimental tests of the Equivalence Principle (EP) are of seminal relevance as probes of General Relativity. The Equivalence Principle is tested by observing its consequence, namely the Universality of Free Fall, whereby in a gravitational field all bodies fall with the same acceleration regardless of their mass and composition.  They therefore require two test masses of different composition, falling in the field of another `source' mass, and a read-out system to detect their motions relative to one another.  An EP violation  would result in a differential displacement of the masses in the direction of the source mass, which cannot be explained on the basis of known, classical phenomena ({\it e.g.} tidal effects).

The landmark  experiment by E\"otv\"os and collaborators~\cite{eotvos} has established that a torsion balance is most well suited for ground tests of the EP, thanks to its inherently differential nature. With the test masses suspended on a torsion balance they improved previous pendulum experiments by almost $4$ orders of magnitude, showing no violation for $\eta$ larger than a few $10^{-9}$~\cite{eotvos}. Several decades later, by exploiting the $24$-hr modulation of the signal in the gravitational field of the Sun, torsion balance tests have improved to $10^{-11}$~\cite{Dicke}  and then to $10^{-12}$~\cite{Bra}. More recently, systematic and very careful tests carried out by Adelberger and co-workers using {\it rotating} torsion balances have provided even more firm evidence that no violation occurs to the level of $10^{-12}$~\cite{Su, Adel}.  

The relevant theoretical question for Equivalence Principle tests is: at which accuracy level a violation, if any,  is to be expected? In an earlier work by Damour and Polyakov, based on string theory and the existence of the dilaton~\cite{Damour1}, $\eta$ values at which a violation might be observed have been determined to be in the range $10^{-18}<\eta<10^{-13}$. Fischbach and coworkers~\cite{Fisch} have derived a non-perturbative rigorous result, according to which a violation must occur at the level of $\eta\simeq 10^{-17}$, due to the coupling between gravity and processes of $\nu-\bar{\nu}$ exchange which should differently affect masses with different nuclei. More recent work~\cite{Damour2} suggests,  in a new theoretical framework for the dilaton, that a violation might occur already at the level of $\eta\simeq 10^{-12}-10^{-13}$, depending on the composition of the masses.

While an $\eta\simeq 10^{-13}$, and perhaps smaller, should be accessible with rotating torsion balance experiments on the ground, a sensitivity as high as $\eta\simeq 10^{-17}$ could be achieved only by an experiment flying in low Earth orbit, where the driving acceleration is up to three orders of magnitude larger. Specific instruments have been designed to carry out such an experiment in space: STEP, Microscope and  ``Galileo Galilei'' (GG)~\cite{STEP}-\cite{varenna}. They share two features: that the test masses are concentric cylinders, and that rotation of the spacecraft provides signal modulation at frequencies higher than the orbital one.   

GG is peculiar in that it  spins around the symmetry axis and is sensitive to relative displacements in the plane perpendicular to it:  cylindrical symmetry of the whole system and  rotation around the symmetry axis allow passive attitude stabilization of the spacecraft with no need of a motor after initial spin  up to the nominal frequency (typically $2$ Hz). The planar (instead of linear) sensitivity of the instrument is also a crucial feature for allowing us to rotate at supercritical speeds, {\it i.e.}  faster than the natural frequencies of the system. Faster rotation means modulation of the signal at higher frequency and therefore a reduced $1/f$ noise (for $1/f$ noise see {\it e.g.} the website maintained by W. Li~\cite{unosuf}). GG differs from the other proposed space experiments also in that the test masses are suspended mechanically. We find that in absence of weight, as it is the case in space, mechanical suspensions too can provide extremely weak coupling, with the additional  advantage to electrically ground the test masses. 

The GG design naturally allows us to build and test a full scale $1$-g version of the apparatus: by suspending the instrument on a rotating platform through its spin/symmetry axis, the sensitive plane lies in the horizontal plane of the laboratory where a component of an EP violation signal might be detected, similarly to a torsion balance experiment.  ``Galileo Galilei on the Ground'' (GGG)~\cite{GGG1,GGG2} is primarily a prototype for testing the main novel features of the experiment proposed for flight. It is also an EP experiment in its own right aiming to compete with  torsion balance tests~\cite{Su, Adel}. In this effort,  motor noise, low frequency terrain tilts~\cite{Moriond} and tidal perturbations~\cite{Raffa} are the  main issues to be addressed.

A full knowledge of the dynamical response of the GGG rotor is needed, especially in view of its condition of supercritical rotation, and of its common mode rejection behavior. The theoretical understanding of the dynamical properties of the rotor, together with the construction of a full simulation facility, would allow us to predict and interpret the collected experimental data; they also provide a virtual environment for planning the experiment and optimizing its performance.

With these motivations in mind, we demonstrate that a simple but very effective mathematical model can be set up to quantitatively describe the dynamical properties of the GGG rotor. In this paper (Part I), we determine the normal modes in all regimes, from subcritical to supercritical rotation, and address the issue of self-centering in supercritical rotation. In the following paper (Part II), we provide the dependence of the common mode rejection ratio on various system parameters which govern the design of the instrument.

The differential equations in the model are solved by means of a user-friendly simulation program and the numerical solutions are tested against the data available from the experiment. The physical content of the model is also discussed by means of approximate analytical solutions, which provide useful physical insight.

This paper is organized as follows: Sec.~\ref{SecExp} describes the main features of the experimental apparatus; Sec.~\ref{SecMod} presents the dynamical model of the system, referring to specific appendices for  details. Sec.~\ref{SecNum} reports on  the numerical method that we have implemented and Sec.~\ref{SecResNormal} gives  the results obtained on the determination of the normal modes of the system, showing an excellent agreement between theoretical predictions and experimental data. The details of the calculations are contained in two appendices, while the third one is specifically devoted to the important concept of self-centering. Concluding remarks are discussed in Sec.~\ref{SecCon}.

\section{The GGG rotor: overview of the experiment}\label{SecExp}

GGG is a rotating differential accelerometer operated in a vacuum chamber (see Fig.~\ref{Fig:GGGscheme}). It is made of two concentric hollow test cylinders, $10$ kg each, weakly coupled by means of a vertical arm $-$a tube located along the axis of the cylinders$-$ to form a vertical beam balance (from now on we shall always omit the term ``hollow'' when
referring to the test cylinders). The coupling arm is suspended at its midpoint from a rotating vertical shaft in the shape of a tube enclosing it (see Fig.~\ref{Fig:GGGscheme} and Fig.~\ref{Fig:photos}, right hand side). A total of three suspensions are needed (drawn in red in Fig.~\ref{Fig:GGGscheme}): a central one (see Fig.~\ref{Fig:photos}, left hand side) to suspend the coupling arm from the rotating shaft, and one for each test cylinder to suspend each of them from the top and bottom end  of the vertical coupling arm. 

The suspensions are cardanic laminar suspensions manufactured in CuBe which are stiff in the axial direction $\hat{Z}$, against local gravity, and soft in the plane $\hat{X}-\hat{Y}$ orthogonal to the axis so that the geometry is naturally two-dimensional, the horizontal plane being sensitive to differential accelerations acting between the test cylinders. In normal operation mode, modulation of such a signal is provided by setting the whole system in  rotation around the vertical axis in {\it supercritical regime}, namely at frequencies $\nu_s$ larger than the natural differential frequencies of the rotor, typically $\nu_s>1.5$ Hz. The differential character of the instrument is strengthened by two differential read-out systems made of four capacitance plates (indicated as $IP$, internal plates, in Fig.~\ref{Fig:GGGscheme}) located in between the  test cylinders and which are part of two capacitance bridges in two orthogonal directions of the sensitive plane.

\subsection{Description of the mechanical structure of the apparatus}

\begin{figure}[htbp]
\includegraphics[width=3.2in]{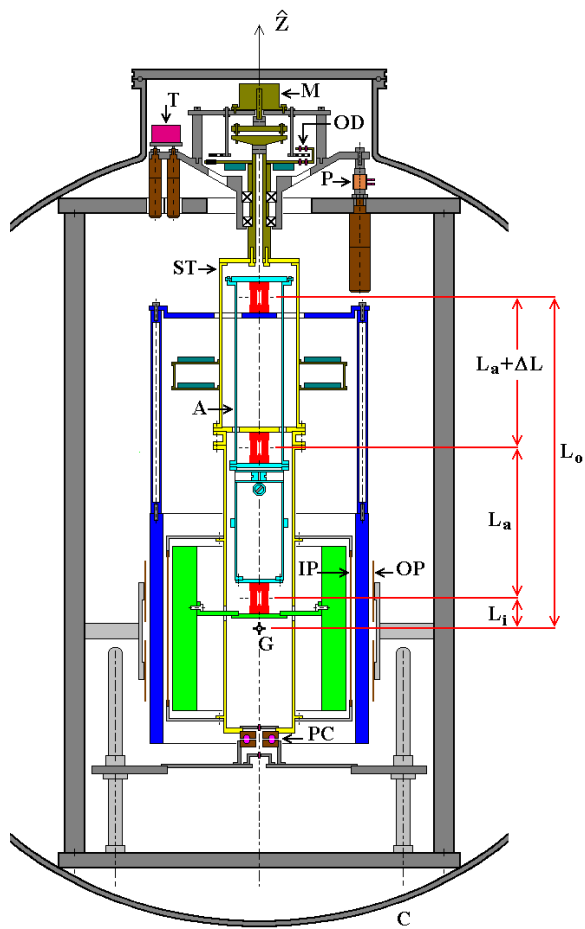}
\caption{Section through the spin axis $\hat{Z}$ of the differential accelerometer inside the vacuum chamber. $C$: vacuum chamber; $M$: motor; $OD$: optical device (see Sec.~\ref{Secsecread}); {\tt x}: ball bearings; $ST$: suspension tube; $A$: coupling (balance) arm, located inside the suspension tube, with its $3$ laminar cardanic suspensions (in red); $G$: center of mass of the two cylinder's system (in blue the outer cylinder, in green the inner one, $10$ kg each). $IP$ are the internal capacitance plates of the differential motion detector (Sec.~\ref{Secsecread}), $OP$ are the outer ones for whirl control (Sec.~\ref{whirl}), $PC$ is the contactless inductive power coupler providing power to the electronics inside the rotor. The relevant distances $L_i$, and $L_o$ of the centers of mass of the inner and outer bodies from their suspension points are also sketched, along with the arm length $2L_a+\Delta L$. $T$ and $P$, at the top of the rotor, are the tiltmeter and $3$ PZTs (at $120^{o}$ from one another -only one shown) for automated control of low frequency terrain tilts. The drawing is to scale and the inner diameter of the vacuum chamber is $1$ m. } 
\label{Fig:GGGscheme} 
\end{figure}

\begin{figure}[htbp] 
\includegraphics[width=3.42in]{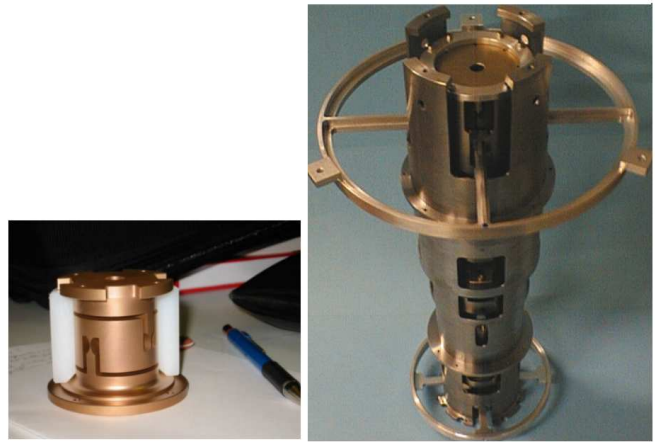}
\caption{Left hand side: the central laminar cardanic suspension of the GGG rotor, located at the midpoint  to  the coupling arm in order to suspend it from the suspension tube (shaft). Right hand side: the coupling arm inside the suspension tube (shaft) as seen from the top. Two cardanic laminar suspensions are located at its top and bottom ends. They suspend the  test cylinders (not shown here) through two metal rings. The dimensions of the rings depend on the dimensions of the concentric  cylinders, which have equal mass ($10$ kg) and therefore different size. Top and bottom ring refer to outer and inner test cylinder respectively.} \label{Fig:photos} \end{figure}

The GGG apparatus is schematically presented in Fig.~\ref{Fig:GGGscheme}, where a section through the spin-symmetry axis $\hat{Z}$ is shown inside the vacuum chamber $C$. At the top-center of the frame is the motor $M$ whose shaft is connected to the  suspension tube of the rotor $ST$ (drawn in yellow) by means of an appropriate motor-rotor joint and turns in the vertical direction inside ball bearings, indicated by {\tt x} symbols in the figure. From the suspension tube $ST$ rotation is then transmitted to a tube located inside it which constitutes the vertical beam of the balance (also referred to as the coupling arm, Fig.~\ref{Fig:photos}, right hand side), the connection between the two being provided at the midpoint of the arm by the central laminar cardanic suspension (see Fig.~\ref{Fig:photos}, left hand side).

The coupling arm in its turn transmits rotation to both the test cylinders, since they are suspended (by means of two laminar cardanic suspensions similar to the central one) from its two ends. More precisely, the  inner test cylinder (shown in green) is suspended from the bottom of the coupling arm at a distance $L_i$ from the cylinder's center-of-mass, while the outer one (shown in blue), is suspended from the top of the coupling arm,  at a distance $L_o$ from the cylinder's center-of-mass.  In Fig.~\ref{Fig:GGGscheme} the three suspensions are drawn in red. It is apparent that the central suspension carries the whole weight of the rotor, mostly the weight of the two test cylinders ($10$ kg each)  plus the small weight of the coupling arm. It is worth noting that the metallic suspensions provide passive electrostatic discharging of the test masses.

In this way, the symmetry  of the whole apparatus is cylindrical, its axis being both the vertical beam of the balance and the axis of  rotation, the balance is sensitive in the horizontal plane and the test masses are concentric.

\subsection{The differential motion detector system}\label{Secsecread}

The Differential Motion Detector (DMD) reflects the cylindrical symmetry of the system and is composed of the following three parts.

(1) Two capacitance plates $IP$ working as DMD(x) sensors (drawn as vertical lines in between the cylinders in Fig.~\ref{Fig:GGGscheme}), are located halfway in between the test cylinders in correspondence to the $X$ direction with a clear gap of $5$ mm on either side and are connected to the suspension tube by means of an insulating frame. A similar pair of capacitances forming a DMD(y) is placed in the $Y$ direction. A voltage signal is applied to each capacitance bridge in order to shift the signal of interest to a high-frequency band with reduced $1/f$ noise (with phase locked detection).  The filtered signal is digitized by an ADC before transmission to the non-rotating (laboratory) frame.  Calibration and balancing of the capacitance bridge are performed by means of procedures outlined in~\cite{GGG1}. The best sensitivity achieved in bench tests corresponds to mechanical displacements of $5$ pm in $1$ s of integration time~\cite{GG,GGG1}. Presently, the sensitivity of the read-out system during normal operation is $\simeq 10^{-9}$ m.

(2) An optical device $OD$ located below the motor and above the ball bearings, utilizing a disk with $32$-holes and an infra-red emitter-detector pair, provides a reference signal for the angular position of the rotor. The reference signal is combined with the $X$ and $Y$ channel data from the DMD and encoded into RS232 format for transmission to a computer. Then a second emitter-detector pair located at the very  bottom of the rotor (using a hole along the axis of the power coupler $PC$, see Fig.~\ref{Fig:GGGscheme}) transmits the digital signal from the rotor to the non-rotating frame from where it is taken out of the vacuum chamber through electrical feed-throughs.

(3) An annular disk, in two semicircular parts, is mounted around the upper half of the suspension tube and contains the two capacitance-bridge circuits and their preamplifiers (see Fig.~\ref{Fig:GGGscheme}). The necessary electronics to demodulate the signal and convert it from analogue to digital form, as well as the drivers for the optical emitter are also located here.

\subsection{Principle of operation}

For detecting an EP violation signal the instrument relies on its sensitivity to relative displacement of the two test masses, which in the final design will be made of different materials.  An acceleration   in the horizontal plane of the laboratory acting differently between the test cylinders gives rise to a relative displacement of the two in the direction of the acceleration. This displacement unbalances the capacitance bridges and gives rise to an electric voltage proportional to it. 

A modulation of the displacement, as seen by the capacitance plates, is achieved by setting the whole system in rotation around the vertical axis of symmetry passing through the shaft, as shown in Fig.~\ref{Fig:GGGscheme}.  Note that the signal modulation obtained in this way does not affect the centers of mass of the test cylinders, hence it does not affect  their relative displacement, which is the physical quantity measured in the experiment. As a result, this type of modulation reduces the noise but not the signal. 

In point of fact, this signal  modulation could be achieved by keeping the test cylinders stationary and rotating only the capacitance plates (located in between the two, indicated as $IP$ in Fig.~\ref{Fig:GGGscheme}) which form the differential motion detector system described above. However, by rotating the test cylinders together with the capacitors, any irregularity in their mass distribution averages out; moreover, the supercritical regime can be exploited to reduce rotation  noise for all parts of the apparatus (see Sec.~\ref{whirl} and Appendix~\ref{SecResSC}). As for the experiment in space, rotation of the whole spacecraft has two more very important advantages. In the first place, it eliminates the need for motor and ball bearings altogether, which are a considerable source of noise in the ground experiment. Secondly, by rotating around the axis of maximum moment of inertia, the spacecraft is {\it passively} stabilized, thus reducing its weight, cost and complexity, as well as disturbances on the EP experiment.

An EP violation signal in the gravitational field of either the Earth or the Sun would have a component in the horizontal plane of the laboratory which could  be detected by the instrument. Since the test bodies are rotors suspended on the Earth, and the Earth rotates around its axis, this diurnal rotation gives rise to large gyroscopic effects on the test bodies resulting in a non zero differential acceleration which would mask an  EP violation signal in the field of the Earth itself. Measurements of such gyroscopic effects have been  reported in~\cite{GGG1} (Sec. 5, Fig. 12). The instrument $\--$in this ground based version$\--$ is therefore  used  for two purposes: {\it i)} to establish its sensitivity as a prototype of the flight instrument, namely for an expected signal at the orbital frequency of the satellite ($\simeq 1.75\cdot10^{-4}$ Hz, {\it i.e.} about  $1$ and half hour period, at an Earth orbiting altitude of $\simeq 520$ km); {\it ii)} to look for  an EP violation in the gravitational field of the Sun, in which case the signature of the signal (see ~\cite{varenna}, Sec. 2) would  have a dominant Fourier component of $24$-hr period due to the diurnal rotation of the Earth.

\subsubsection{Differential character and common mode rejection}

The differential character of the whole instrument, namely its capability to reject accelerations which are common to both test masses, is in principle ensured by the geometry and mounting of the test masses.  It is further augmented by the differential nature of the DMD system.

{\it (i)} The sensitivity of the instrument to differential accelerations of the test masses depends on the softness of the laminar suspensions and on the uniform distribution of mass around the spin axis.  Soft suspensions and a good balancing of the rotor provide long natural periods for differential oscillations of the test masses relative to each other, giving rise to larger relative displacements between the two, and in turn to stronger output voltage signals.

Tuning of the natural differential period $T_D$ of the test cylinders is made possible by changing a moment-arm in the beam balance. This is accomplished by moving a small solid ring mounted at the lower end of the balance (coupling) arm.  Moving this ring vertically along the arm, in the $\hat{Z}$ direction, displaces the center-of-mass of the balance arm from its suspension point by a quantity $\Delta L$. If $\Delta L=0$ the center-of-mass of the balance arm is coincident with its suspension point. $\Delta L$ can be adjusted to be either slightly positive or negative, resulting in a longer or shorter $T_D$. However, there's a maximum positive value that $\Delta L$ can assume before the system becomes unstable (see  Eq. (\ref{eqTdiff}) below).

Asymmetric distribution of mass of the rotor in the horizontal plane, resulting in a non-zero inclination of the coupling arm in the rotating reference frame, may also be corrected by two small masses mounted inside the coupling arm itself, one of which is movable in the X direction and the other in the Y.  

The tilt of the spin axis with respect to the non rotating laboratory frame is controlled by three micrometer screws which support the plate on which the rotor shaft is mounted.  In addition the tilt can be finely adjusted using piezoelectric actuators (P) attached to the tips of the micrometer screws (see Fig.~\ref{Fig:GGGscheme}). 

{\it (ii)} As to the DMD system, a non-zero off-centering of the capacitor plates $IP$ located in between the test cylinders $\--$measured by the ratio  $(a-b)/a$ where $a$ ($b$) is the nominal gap between the inner (outer) mass and any one of the capacitance plates$\--$  would make a common mode displacement $\Delta x_C$ of the test masses to produce  a differential output signal in addition to that produced by a real differential displacement $\Delta x_D$. The larger this off-centering, the larger  the fraction of common mode displacement which is turned into a `fake' differential signal, i.e. which contributes to the total unbalance $\Delta C$ of the capacitance bridge~\cite{GG} (GG Phase A Report, Sec. 2.1.3) from the original capacitance value $C_o$: 
\begin{equation} \frac{\Delta C}{2C_0}\simeq \frac{a-b}{a^2}\Delta x_C-\frac{1}{a}\Delta x_D\; . \end{equation}

\subsubsection{Signal modulation and whirl motions}\label{whirl}

Signal modulation in testing the equivalence principle has been first proposed  in~\cite{Dicke} in order to improve the E\"otv\"os experiments. By referring   to the Sun rather than the Earth as the source mass of the gravitational field, the diurnal rotation of the Earth itself on which the test masses are suspended provides a $24$-hr modulation with no need to rotate the experimental apparatus, an operation which gives rise to relevant disturbances  in such small force experiments. However, higher modulation frequencies are desirable in order to reduce $1/f$ noise, and in fact excellent results have been obtained by~\cite{Su,Adel} with torsion balances placed on a turntable rotating faster than the Earth. In GGG we try to spin the test masses much faster,  at frequencies (typically a few Hz) higher than the natural frequencies $\nu_n$ of the system, a condition known as {\it supercritical rotation}. 

The GGG apparatus has $3$ natural frequencies. The differential frequency $\nu_D$, of the oscillations of the test bodies relative to one another,  and two  common mode frequencies, $\nu_{C1}$ and $\nu_{C2}$, of both test masses together. In the GGG setting reported here their values are: $\nu_D=0.09$ Hz, $\nu_{C1}= 0.91$ Hz, $\nu_{C2}= 1.26$ Hz. 

It is well known~\cite{DenHartog,Crandall,Genta} that in supercritical rotation the masses are able to self-center and greatly reduce the original offsets of their centers of mass with respect to their own rotation axes.  Any initial offset, which inevitably results from construction and mounting errors, is in fact reduced by a factor  $(\nu_D/\nu_s)^2$.  Such self-centering is a very essential requirement when using fast rotating macroscopic test bodies for the purpose of detecting the effects of extremely small forces between them. 

It is also well known that in supercritical rotation, dissipation in the system gives rise to destabilizing {\it whirl} motions at frequencies $\nu_w$ equal (or close) to the natural frequencies of the system, whose amplitude increases with time at a rate $1/\tau_w=\pi \nu_w/{\mathcal Q}(\nu_s)$ scaling as the whirl frequency $\nu_w$ and the inverse of the quality factor ${\mathcal Q}$ at the spin frequency $\nu_s$~\cite{Genta,CrandallNobili,Nobili}.

Whirls can be stabilized by passive and active methods. Passive stabilization is typically used in engineering applications of supercritical rotors, but it produces too large disturbances for our purposes. We have used a passive damper in the past only to stabilize the rotor during resonance crossing (see ~\cite{GGG1}, Sec. $3$). With the current improved apparatus, damping at resonance crossing  is no longer needed. A much finer whirl stabilization can be performed actively by means of $8$ small capacitance sensors/actuators (indicated as $OP$, outer plates, in Fig.~\ref{Fig:GGGscheme}) placed close to the outside surface of the outer test cylinder, $4$ of them used as sensors and $4$ as actuators in two orthogonal directions of the horizontal plane~\cite{GGG2}. In the GGG experiment performed at supercritical speed the relevant ${\mathcal Q}$ value is determined by losses due to deformations of the laminar suspensions at the spin frequency. Experimental  measurements of ${\mathcal Q}$ are reported in ~\cite{GGG1} and, more recently, in ~\cite{GGG2005}

\section{The model}\label{SecMod} 

\begin{figure}[htbp] \hbox{\includegraphics[width=1.7in]{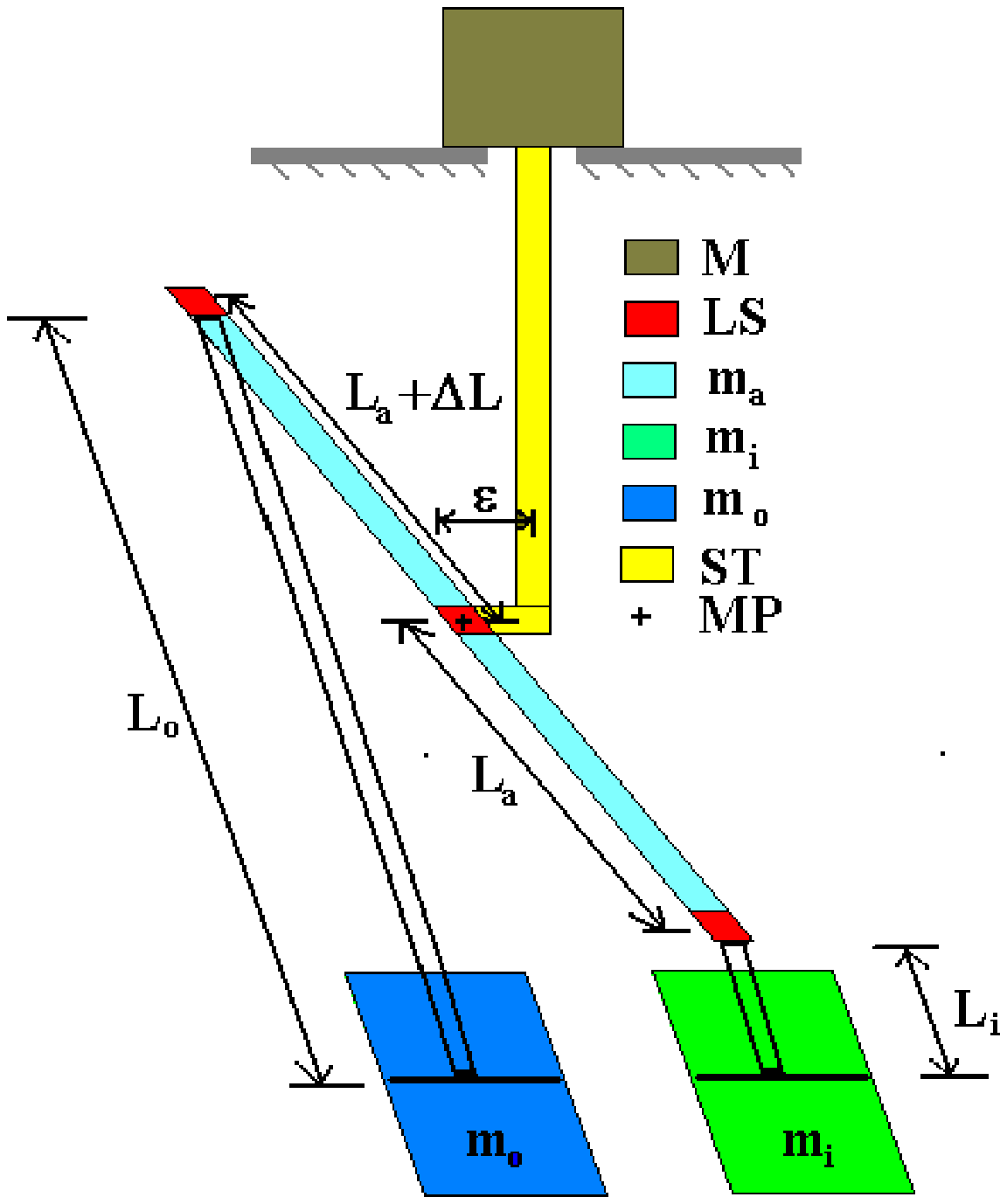} \includegraphics[width=1.7in]{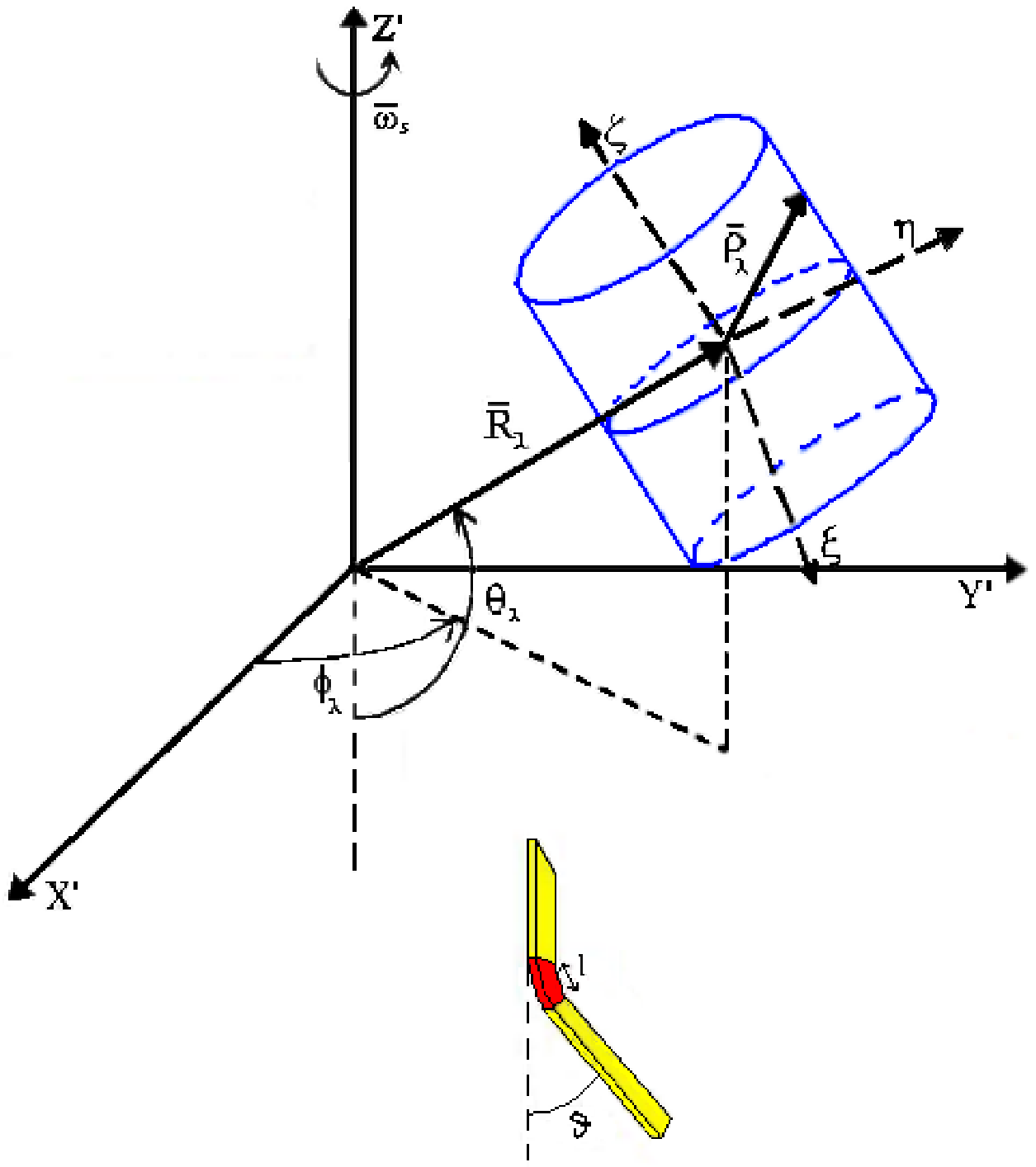}} \caption{Minimal model for the real instrument sketched in Fig.~\ref{Fig:GGGscheme} (see text for details). On the left hand side the various parts are drawn with the same colors and  labels as  in Fig.~\ref{Fig:GGGscheme}. Here the midpoint of the coupling arm is indicated  as $MP$.  $L_a$, $L_o$ and $L_i$ refer to the dimensions of the coupling arm and the  outer mass and inner mass suspension arms respectively. $\hat{L}_a$, $\hat{L}_o$, and $\hat{L}_i$ are the unit vectors of the corresponding beams. The offset vector $\vec{\epsilon}$, due to construction and mounting imperfections, is also indicated. On the right  hand side we sketch one of the cylinders in the rotating reference frame $\{X'Y'Z'\}$, showing its principal axes of inertia $\{\xi,\eta,\zeta\}$, the position vector $\vec{R}_\lambda$ ($\lambda=a,o,i$) of its center of mass and the angles $\theta_\lambda,\phi_\lambda$, which are not the usual Euler angles, as discussed in the text. Below this figure, the small one to the right shows a typical deformation of one of the laminar suspensions of length $l$, for instance the central one. None of these figures is to scale.} \label{fig:model} \end{figure}

Having described the real instrument, we are now in a position to outline the minimal model used to describe its dynamical behavior. Fig.~\ref{fig:model} displays a schematic representation of the model in the reference frame $\{X'Y'Z'\}$ rotating with the shaft at an angular velocity $\omega_s=2\pi\nu_s$ around the $Z'$ axis ($\vec{\omega}_s=\omega_s\hat{Z'}$). The relevant parts of the instrument depicted in Fig.~\ref{Fig:GGGscheme} are sketched in Fig.~\ref{fig:model} with the same colors. The coupling arm, with mass $m_a$ (drawn in cyan as in Fig.~\ref{Fig:GGGscheme}), and length $2L_a+\Delta L$ is suspended at its midpoint $MP$ from the rotating shaft and suspension tube $ST$ (yellow) by means of the central laminar suspension $LS$ (red) with elastic constant $K$. The vector $\vec{\epsilon}$ is the offset of the arm center-of-mass from the axis, which is unavoidable because of construction and mounting errors. Variations of $\Delta L$, as we have already discussed, produce a change of the mass distribution, hence of the natural differential period of the test masses, $T_D$. Here, and with no loss of generality, $\vec{\epsilon}$ is placed along the $X'$ axis.

The outer test cylinder, of mass $m_o$ (blue), is suspended from the top of the coupling arm by means of the laminar suspension with elastic constant $K_o$ and its center of mass is at a distance $L_o$ from the suspension. In a similar manner, the inner test mass $m_i$ (green) is suspended from the bottom  of the arm, $K_i$ and $L_i$ being the corresponding parameters. From now on, the label $\lambda=i,o,a$ will be used to refer to the parameters of the inner mass, outer mass, and coupling-arm respectively. The three bodies have moments of inertia $I_{\lambda\xi}=I_{\lambda\eta}= m_\lambda(3 R_{\lambda I}^2 + 3 R_{\lambda E}^2 + R_{\lambda H}^2)/12$ and $I_{\lambda\zeta}=m_{\lambda}(R_{\lambda I}^2 + R_{\lambda E}^2)/2$ along their principal axes ($\xi$, $\eta$ and $\zeta$),  $R_{\lambda I}$, $R_{\lambda E}$ being the  internal and external radii of the cylinder $\lambda$, and $R_{\lambda H}$ its height.

The laminar suspensions have length $l$,  the central one is slightly stiffer than the other two and we assume  $K_i=K_o\neq K$. In a refined version of the model, and whenever specificed, we also consider an anisotropic central suspension by introducing the parameter $\Lambda$ such that $K_{Y'}\equiv\Lambda K_{X'}$.

By defining the unit vector  $\hat{L}_a$ of the coupling arm as pointing from its midpoint towards the bottom suspension, and the unit vectors $\hat{L}_o$ and $\hat{L}_i$ of the test cylinders each pointing from the suspension to the center of mass of the body (see Fig.~\ref{fig:model}), the corresponding position vectors in the rotating reference frame $\{X'Y'Z'\}$ of Fig.~\ref{fig:model} are  \begin{eqnarray} \vec{R}_a&=&\vec{\epsilon}-0.5\Delta L\hat{L}_a\nonumber\\ \vec{R}_o&=&\vec{\epsilon}-(L_a+\Delta L)\hat{L}_a+L_o\hat{L}_o\label{eqVect}\\ \vec{R}_i&=&\vec{\epsilon}+L_a\hat{L}_a+L_i\hat{L}_i\nonumber\; . \end{eqnarray}

\subsection{The Lagrangean}

The Lagrangean $\cal{L}$ in the rotating reference frame $\{X'Y'Z'\}$ can 
be written as 
\begin{equation}
{\cal L} ={\cal T}-{\cal U}\; ,
\label{eqL}
\end{equation}
where the kinetic term can be very generally written as
\begin{equation}
{\cal T}=\frac{1}{2}
\sum_{\lambda=a,o,i}\int_{\tau_\lambda}\vec{v}_{\lambda}^2
dm_\lambda\; 
\label{eqTgen}
\end{equation}
after defining the velocity $\vec{v}_\lambda$ of the mass element 
$dm_\lambda$ in body $\lambda$ with volume
$\tau_\lambda$. Then, ${\cal U}$ includes 
the potential energies associated with  gravity and with the elastic forces, namely 
\begin{equation}
{\cal U}=U_g+U_{el}\label{eqUgen}
\end{equation}
where 
\begin{eqnarray}
U_g&=&\sum_{\lambda=a,o,i}-m_\lambda\vec{g}\cdot\vec{R}_\lambda
\quad \hbox{\rm with}\quad \vec{g}\equiv -g\hat{Z'}\label{eqUg}\\
U_{el}&=&
\sum_{\lambda=o,i}\frac{1}{2}K_\lambda l^2|\hat{R_a}\times\hat{R}_\lambda|^2
\label{eqUel}\\
&+&
\frac{1}{2}L_a^2\left(K_{X'}|\hat{R_a}\dot\hat{X'}|^2+K_{Y'}|\hat{R_a}
\dot\hat{Y'}|^2\right)
\label{eqU}\; .
\end{eqnarray}
For the expression of $U_{el}$ we refer to the small figure at the bottom right of Fig.~\ref{fig:model},
sketching the laminar suspension and its orientation.

We proceed along the main steps to derive the operational expression for ${\cal L}$. The bodies are rotating around their own axis with angular velocity $\omega_s$ in a reference frame which is rotating as well, as sketched in Fig.~\ref{fig:model}, right hand side. Thus, we define as $\vec{\Omega}_\lambda$ the angular-velocity vector of the element $dm_\lambda$ in the $\{X'Y'Z'\}$ frame and $\vec{\omega}_s=\omega_s\hat{Z}'$, so that 
\begin{equation} \vec{v}_\lambda=\vec{V}_\lambda+\vec{\Omega}_\lambda\times\vec{r}_\lambda = \vec{V}_\lambda+\vec{\Omega}_\lambda\times(\vec{R}_\lambda+\vec{\rho}_\lambda) \label{eqv}\; , 
\end{equation} 
where $\vec{V}_\lambda$ is the velocity of the center-of-mass of body $\lambda$ and $\vec{r}_\lambda$ is the vector pointing to the element $dm_\lambda$, composed by $\vec{R}_\lambda$ and $\vec{\rho}_\lambda$ as drawn in Fig.~\ref{fig:model}. By inserting Eqs.(\ref{eqv}) into Eq.~(\ref{eqTgen}), we can write ${\cal T}$ as 
\begin{equation} {\cal T}=T_{kin}+T_{Cor}+U_{Cor}+U_{c}\; , \label{eqT} 
\end{equation} 
where the only non-zero terms are (see Appendix~\ref{appA} for details) 
\begin{eqnarray} T_{kin}&=&\frac{1}{2}\sum_\lambda\left(m_\lambda V_\lambda^2+ \sum_{\alpha=\xi,\eta,\zeta} I_{\lambda\alpha\alpha}\Omega_\alpha^2\right)\label{eqTkin}\\ T_{Cor}+U_{Cor}&=&\sum_\lambda m_\lambda \vec{V}_\lambda\cdot\left(\vec{\omega_s}\times\vec{R}_\lambda\right) \label{eqTCor}\\ &+&\sum_\lambda\int_{\tau_\lambda}\left(\vec{\Omega}_\lambda\times \vec{\rho}_\lambda\right) \cdot\left(\vec{\omega}_s\times\vec{\rho}_\lambda\right)dm_\lambda\label{eqUCor}\\ U_c&=&\frac{1}{2}\sum_\lambda\int_{\tau_\lambda} \left[\vec{\omega_s}\times \left(\vec{R}_\lambda+\vec{\rho}_\lambda\right)\right]^2dm_\lambda \label{eqCen}\; . 
\end{eqnarray}

In Eqs.~(\ref{eqTCor})-(\ref{eqUCor}) the terms coming from Coriolis forces have been split into the $U_{Cor}$ potential energy, which contains only the position vectors, and $T_{Cor}$ which contains also the velocities. The centrifugal part $U_c$ has been indicated as a potential energy. To proceed further, we now have to specify the choice of the generalized coordinates.

\subsection{Choice of the generalized coordinates}\label{SecsecGenCoo}

The GGG rotor model shown in Fig.~\ref{fig:model} is composed of $n_b=3$ coupled bodies, for a total of $18$ degrees of freedom. However, the  central suspension prevents them from performing translational motions, thereby reducing the degrees of freedom to $9$. In addition, the motor forces the three bodies to rotate at a constant angular velocity, so that the number of degrees of freedom for  the model is  $n=6$.

We have chosen as generalized coordinates for each body the two angles $\theta_\lambda$ and $\phi_\lambda$ (see Fig.~\ref{fig:model}, right hand side). These angles are defined  slightly differently from the usual Euler angles: $\theta_\lambda$ is the angle between $\vec{R}_\lambda$ and the axis $-\hat{Z}'$ and runs in the interval $[0,\pi]$;  $\phi_\lambda$ is the angle from the $\hat{X}'$ axis to the projection of $\vec{R}_\lambda$ on the $X'Y'$ plane and runs in the interval $[0,2\pi]$. We thus define the vector $Q$ of the generalized coordinates and the corresponding velocities $\dot{Q}$
\begin{equation}
Q=\{q_1,q_2,q_3,q_4,q_5,q_6\}=
\{\theta_a,\phi_a,\theta_o,\phi_o,\theta_i,\phi_i\}\; .
\label{genQ}
\end{equation}

With these definitions in hand, we have that 
\begin{equation}
\hat{L}_a  = \{\sin\theta_a\cos\phi_a,
\sin\theta_a\sin\phi_a,-\cos\theta_a\}\; ,
\label{eqVers}
\end{equation}
and similar expressions
for $\hat{L}_o$ and $\hat{L}_i$. Eq.~(\ref{eqVers})
turns Eqs. (\ref{eqVect}) into 
expressions for the 
$\vec{R}_\lambda(Q)$  and the corresponding velocities 
$\vec{V}_\lambda=\dot{\vec{R}}_\lambda(Q,\dot{Q})$. We then conveniently 
write all the
vectors in the $\{X'Y'Z'\}$ reference frame in terms of their components in the 
$\{\xi\eta\zeta\}$ frame by means of the rotation matrix ${\cal M}$ 
(Eq.~\ref{eqMRot}), 
namely $\vec{\Omega}_\lambda=\stackrel{\leftrightarrow}{{\cal M}}
\vec{\Omega}_{\lambda,\xi\eta\zeta}$ 
and $\vec{\rho}_\lambda=\stackrel{\leftrightarrow}{{\cal M}}
\vec{\rho}_{\lambda,\xi\eta\zeta}$. 

After noting that  
$\vec{\Omega}_{\lambda,\xi\eta\zeta}=\{-\dot{\theta}_\lambda,
\dot{\phi}_\lambda\sin\theta_\lambda,\omega_s\}$ and performing all the 
integrals over the three bodies, we finally obtain (Appendix~\ref{appA})
the operative expression for ${\cal L}(Q,\dot{Q})$ in the rotating 
reference frame:
\begin{equation}
{\cal L}(Q,\dot{Q})=T(Q,\dot{Q})-U(Q)\; ,
\label{eqLdef}
\end{equation}
where we have defined 
\begin{eqnarray}
T(Q,\dot{Q})\equiv T_{kin}(Q,\dot{Q})+T_{Cor}(Q,\dot{Q})\label{eqTtot}\\
U(Q)\equiv U_g(Q)+U_{el}(Q)-U_{Cor}(Q)-U_c(Q)\; .
\label{eqUtot}
\end{eqnarray}
The terms entering (\ref{eqLdef})-(\ref{eqUtot}) are
\begin{eqnarray}
T_{kin}&=&\frac{1}{2}\sum_\lambda\left[m_\lambda V_\lambda(Q,\dot{Q})^2+
I_{\lambda\xi}\left(\dot{\phi}_\lambda^2\sin^2\theta_\lambda+
\dot{\theta}_\lambda^2\right)\right]\; ,
\nonumber\\
T_{Cor}&=&\sum_\lambda m_\lambda
\vec{V}_\lambda(Q,\dot{Q})\cdot\left(\vec{\omega_s}\times\vec{R}_\lambda(Q)\right)
\\
&&+\sum_\lambda I_{\lambda\xi}\omega_s\dot{\phi}_\lambda\sin^2\theta_\lambda\; ,
\nonumber\\
U_{Cor}&=&-\sum_\lambda I_{\lambda\zeta}\omega_s^2\cos\theta_\lambda\nonumber
\label{eqUCoriolis}\; ,
\end{eqnarray}
and
\begin{eqnarray}
U_c&=&\frac{1}{2}\sum_\lambda m_\lambda\left(\vec{\omega}_s\times
\vec{R}_\lambda(Q) \right)^2\nonumber\\
&+&\frac{1}{2}\sum_{\lambda}
\left[I_{\lambda\xi}\sin^2\theta_\lambda+
I_{\lambda\zeta}\cos^2\theta_\lambda\right]\omega_s^2\label{eqCenden}\; .
\end{eqnarray}
To these equations we have to add the expressions (\ref{eqUg})-(\ref{eqUel}) written 
in terms of $\vec{R}_\lambda(Q)$ 
through (\ref{eqVect}) and (\ref{eqVers}).

Eqs.~(\ref{eqLdef}) together with  
Eqs.~(19)-(\ref{eqCenden}) yield 
the Lagrange function of the model in Fig.~\ref{fig:model}.

\subsection{Equilibrium positions and second-order expansion}

During normal and successful operation of the GGG rotor only very small 
amplitude motions take place. The Lagrange function (\ref{eqLdef}) can thus be
expanded to second order in $(Q,\dot{Q})$ around the equilibrium solution
$(Q^0,\dot{Q}=0)$, $Q^0=\{q_1^0,...,q_6^0\}$ to derive linearized equations 
of motion. 

In order to do this, we first determine the equilibrium positions from the equation
\begin{equation}
\frac{\partial U}{\partial q_j}|_{q_j=q_j^0}=0\qquad 
j=1,...,n
\; .
\label{eqMin}
\end{equation}

We then use the physical assumption that during the motion, the $Q$'s are 
sligthly perturbed from their equilibrium values $Q^0$. This results in 
the substitutions
\begin{eqnarray}
Q&\rightarrow& Q^0+Q\label{eqLinQ}\\
\dot{Q}&\rightarrow&\dot{Q}\label{eqLinP}\; .
\end{eqnarray}
into (\ref{eqLdef}) to obtain 
a linearized version of the Lagrange function. ${\cal L}(Q,\dot{Q})$ can 
now be expanded to 
second order, namely
\begin{eqnarray}
{\cal L}(Q,\dot{Q})&=&a_0+\sum_{j<k}^n a_{jk}q_j q_k+
\sum_{j<k}^n b_{jk}\dot{q}_j \dot{q}_k+
\sum_{j,k=1}^n c_{jk}q_j \dot{q}_k\nonumber\\
&+&\sum_{j=1}^n d_{j}\dot{q}_j+O(q_j,\dot{q}_k)^4\; ,
\label{eqLexp}
\end{eqnarray} 
where we remark that now the $q_j$'s are small according to the substitutions
(\ref{eqLinQ})-(\ref{eqLinP}), and that the linear terms have cancelled out 
because of
(\ref{eqMin}). The matrix coefficients $a_{jk}$, $b_{jk}$, 
and $c_{jk}$ are known functions of the $Q^0$ and of the governing parameters 
of the system, and in general are to be numerically evaluated.

\subsection{Linearized equations of motion}

The equations of motion in terms of the known $a_{jk}$, $b_{jk}$, 
and $c_{jk}$ coefficients are 
\begin{equation}
\frac{d}{dt}\frac{\partial {\cal L}}{\partial\dot{q}_j}-
\frac{\partial {\cal L}}{\partial q_j}={\cal F}_j\; ,\qquad j=1,...,n
\label{eqMotgen}
\end{equation}
where we have introduced the generalized forces 
\begin{equation}
{\cal F}_j=\sum_{d=1}^3\vec{F}_{\lambda d}\frac{\partial\vec{R}_{\lambda d}}
{\partial q_j}
\label{eqGenF}
\end{equation}
starting from the cartesian components 
$\vec{F}_{\lambda d}$ of the forces acting on each body. 
The ${\cal F}_j$ are to be consistently expanded to first order, namely 
\begin{equation}
{\cal F}_j=\sum_{k=1}^6 \alpha_{jk}q_k+\sum_{k=1}^6 \beta_{jk}\dot{q}_k\; .
\label{eqGenFexp}
\end{equation}

By combining Eqs.~(\ref{eqLexp})-(\ref{eqGenFexp}) together, the equations of motion
can be written in a compact matrix form as
\begin{equation}
\Rmat{M}\ddot{Q}=\Rmat{S}\left(
\begin{array}{l}
Q\\
\dot{Q}\\
\end{array}
\right)\; ,
\label{eqMotC}
\end{equation}
with the obvious notation $\ddot{Q}=\{\ddot{q}_1,...,\ddot{q}_6\}$. In 
Eq.~(\ref{eqMotC}), 
$\Rmat{M}$ is the $n\times n$ ($n=6$) ``mass''-matrix composed by the $b_{jk}$ 
coefficients:
\begin{equation}
\Rmat{M}_{jk}=2b_{jk}\delta_{jk}+b_{jk}(1-\delta_{jk})\label{eqMdef}\; ,
\end{equation}
where the factor of $2$ on the diagonal elements is a consequence of the 
restricted $j<k$
sum in the expansion (\ref{eqLexp}). $\Rmat{S}$ is a $n\times 2n$ matrix 
containing 
the $a_{jk}$, $c_{jk}$, $\alpha_{jk}$, and $\beta_{jk}$ coefficients:
\begin{equation}
\Rmat{S}=\Rmat{A}_2+\Rmat{C}_2+{\Rmat{A}_1}+{\Rmat{B}_1}\; ,
\label{eqSdef}
\end{equation}
with
\begin{eqnarray}
{\Rmat{A}_2}_{jk}={\Rmat{A}_2}_{kj}&=&2a_{jk}\delta_{jk}+a_{jk}
(1-\delta_{jk}) \quad k\leq n\label{eqA2def}\\
          &=&0             \hspace{3.5cm} n<k\leq 2n \nonumber \; , 
\end{eqnarray}
\begin{eqnarray}
{\Rmat{C}_2}_{jk}&=&0\hspace{4.8cm} k\leq n\label{eqC2def}\\
          &=&-{\Rmat{C}_2}_{kj}=c_{jk}-c_{kj}\qquad\qquad\qquad  n<k\leq 
2n \nonumber \; , 
\end{eqnarray}
\begin{eqnarray}
{{\Rmat{A}_1}}_{jk}={{\Rmat{A}_1}}_{kj}&=&\alpha_{jk} \hspace{3.3cm} k
\leq n\label{eqAdef}\\
          &=&0             \hspace{3.5cm} n<k\leq 2n \nonumber \; , 
\end{eqnarray}
and 
\begin{eqnarray}
{{\Rmat{B}_1}}_{jk}&=&0\hspace{4.8cm} k\leq n\label{eqBdef}\\
        &=&{\Rmat{B}_1}_{kj}=\beta_{jk}   \hspace{3.3cm} n<k\leq 2n 
\nonumber \; .
\end{eqnarray}
Note that while $\Rmat{M}$ and the submatrix defined by the first $n=6$ 
columns of $\Rmat{A}_2$ is symmetric, 
the submatrix defined by the second $n=6$ columns of $\Rmat{C}_2$ is 
antisymmetric, as
expected after inspection of the expansion (\ref{eqLexp}).

For all practical purposes, it is convenient to turn (\ref{eqMotC}) into a more
symmetric form involving only first-order time derivatives. To this aim, we
define the $2n=12$-component vector $X$ as
\begin{eqnarray}
X_{2j-1}&=&q_j\nonumber\\
&&\qquad\qquad\qquad j=1,...,n=6\label{eqXdef}\\
X_{2j}&=&\dot{q}_j\nonumber\; .
\end{eqnarray}
By inserting the definition~(\ref{eqXdef}) into (\ref{eqMotC}), we finally obtain
\begin{equation}
\dot{X}=AX\; ,
\label{eqMotD}
\end{equation}
where $A$ is now the square $2n\times 2n$
dynamical matrix defined from $\Rmat{M}^{-1}$ and $\Rmat{S}$ after 
inserting rows of zeros.
\begin{equation}
\begin{array}{llll}
      &=\left(\Rmat{M}^{-1}\Rmat{S}\right)_{j\; 1+\frac{k-1}{2}}&j\; 
{\rm even\; and}\; k\; {\rm odd}\\
A_{jk}&=\left(\Rmat{M}^{-1}\Rmat{S}\right)_{j\; 7+\frac{k-2}{2}}&j\; 
{\rm even\; and}\; k\; {\rm even}\\
      &=1                                         
&j\; {\rm odd\; and}\; k=j+1\\
      &=0                                         
&j\; {\rm odd\; and}\; k\neq j+1\; .\\
\end{array}
\label{eqDdef}
\end{equation}

The relations (\ref{eqMotD})-(\ref{eqDdef}) are central equations, 
written in a form amenable for numerical evaluation. The eigenvalues of the
$A$ matrix (\ref{eqDdef}) correspond to 
the normal modes of the rotor, and the solution of the set 
of differential equations (\ref{eqMotD}) completely determines 
the small-amplitude dynamical behavior of the rotor 
modelled in Fig.~\ref{fig:model}. Before turning to the description of the 
numerical method, 
we introduce rotating and non-rotating damping. 

\subsubsection{Rotating and non-rotating damping}\label{damping}

By means of (\ref{eqGenF})-(\ref{eqGenFexp}) we can in principle introduce any known 
force determining the dynamical behavior of the rotor. In the following we 
 include dissipative forces ${\cal R}_{R}$ and ${\cal R}_{NR}$
due to rotating  and  non-rotating damping
mechanisms respectively (see ~\cite{Crandall,Genta}). The rotating part of the dissipative force 
is to be ascribed to dissipation of the laminar suspensions.
In supercritical rotation, this kind of dissipation is known
to destabilize the system, generating whirl motions. It
can be expressed as 
\begin{eqnarray}
{\cal R}_{R}(Q,\dot{Q})&=&
-\Gamma_{Ra}(\dot{\hat{L}}_a-(\dot{\hat{L}}_a\cdot\hat{Z}')
\cdot\hat{Z}')\label{eqFR}\\
&-&
\sum_{\lambda=o,i}\Gamma_{R\lambda}(\dot{\hat{L}}_\lambda-
(\dot{\hat{L}}_\lambda\cdot\hat{R}_a(Q))\hat{R}_a(Q))\; ,
\nonumber
\end{eqnarray}
where the velocities $\dot{\hat{L}}_\lambda$ are functions of 
$(Q,\dot{Q})$.  Instead, non-rotating damping has  
the effect of stabilizing a system in supercritical 
rotation, and 
can be written as
\begin{equation}
{\cal R}_{NR}(Q,\dot{Q})=-\Gamma_{NR}(\vec{V}_i(Q,\dot{Q})+
\vec{\omega}_s\times\vec{R}_i(Q))\; .
\label{eqFNR}
\end{equation}

Other forces acting on the rotor, such as external disturbances due -for instance- to tides  
 and seismic noise, or control forces applied in order to control the rotor 
dynamics, can also be included,  as described in Sec.~\ref{SecNum}.

\section{The numerical method}\label{SecNum}

\subsection{General considerations} 

The simulation method that we have implemented, rigorously
follows the derivation outlined in Sec.~\ref{SecMod}. We have
found very convenient to use the MATLAB environment, with 
 SYMBOLIC TOOLBOX and SIMULINK packages, 
as it allows us to perform all the needed symbolic calculations
and numerical evaluations, together with the  analysis
of experimental data. 

We start from the formal Lagrange function 
written in a user-friendly way as 
in (\ref{eqL})-(\ref{eqU}) 
and (\ref{eqTkin}-\ref{eqCen}) by means of symbolic vector operations. 
We 
specify the choice (\ref{genQ}) for the generalized coordinates
with respect to the $\{X'Y'Z'\}$ reference frame and define 
accordingly all
the vectors entering ${\cal L}$. We then move on to
the symbolic computation by linearizing and expanding the Lagrange
function as in (\ref{eqLexp}), and define the matrices 
$\Rmat{M}$, $\Rmat{A}_2$, 
$\Rmat{C}_2$, ${\Rmat{A}_1}$, ${\Rmat{B}_1}$, and $A$.  

Once the system parameters are fixed
(see below), the numerical computation is carried out using standard packages to find eigenvalues and eigenvectors
of the $A$ matrix, that are the normal frequencies and modes 
of the spinning rotor. 
The $A$ matrix is then inserted as input to perform the 
dynamical simulation within standard transfer-function method  
used in the SIMULINK toolbox.

The advantage of this strategy is apparent, in that it  easily allows 
us to make any changes in the model that correspond to  changes in the 
experiment we would like to test before implementation. Since  the number of bodies $n_b$ and of the generalized coordinates
$n$ are  symbolically defined and specified only once, all what is to be done
in order to introduce any  changes or new features amounts to modifying or add pieces of the Lagrangean after having symbolically 
written them in terms of vector operations. 

The description of the method used to 
introduce external forces is postponed to Part II of this work, 
where it is used to evaluate the common mode rejection function. We now turn  to
listing the system parameters.

\subsection{System parameters}\label{Secsecsysp}

The parameters which govern the  physics of the GGG rotor are: the geometrical dimensions 
of the three bodies, their weight, the mounting error $\epsilon$, 
the elastic constants, length
and anisotropy factor $\Lambda$  
of the three laminar suspensions, the quality factor ${\mathcal Q}$.
To these parameters $\--$which are fixed after construction$\--$ 
we must add the spin frequency $\nu_s=\omega_s/2\pi$, that can
be varied in the course of the experiment.  
The balancing of the beams and the natural  period $T_D$ of oscillation of the test cylinders relative to one another  
can also be adjusted $\-$as  discussed earlier$\-$  by moving small masses along 
the balance (coupling) arm. 

We have inserted as inputs to the numerical 
calculation all the above parameters as determined in the real GGG instrument. They are
listed in Tab.~\ref{T:1} and~\ref{T:2}. As for the spin frequency, in the experiment it varies 
in the range $(0\leq \nu_s\leq 3.9)\ Hz$ while in the model calculations it can be assumed in a wider range $(0\leq \nu_s\leq 10)\ Hz$. 

The differential period $T_D$ corresponding to the value $\Delta L$ listed 
in Tab.~\ref{T:1} is measured to be $11.7\; s$ and $10.8\; s$ 
in the X and Y directions respectively. These values are in reasonable agreement with 
the following simple  formula: 
\begin{equation}
T_D=\frac{2\pi}{\sqrt{\frac{(K+K_i+K_o)l^2}{(m_i+m_o)L_a^2}
-\frac{g}{2L_a}\frac{\Delta L}{L_a}}}\; 
\label{eqTdiff}
\end{equation}
which can be derived from the general equations of motion (\ref{eqMotgen})
describing  the small oscillations of the $\theta_\lambda$ angles,  
in the very simplified case in which the bodies are neither rotating nor 
subjected to any dissipative or other external forces, 
except gravity, and
under the reasonable assumption that $\theta_i=\theta_o=0$ and that 
$\phi_\lambda$'s are costant, {\it e.g.} $\phi_\lambda=0$.  

Of the whole set of parameters used, only 
the anisotropy factor $\Lambda$ of the suspensions and the construction and
mounting error $\epsilon$ are not measured from the instrument. 
$\Lambda$ is tuned, together with the balancing $\Delta L$, 
so as to reproduce 
the natural frequencies of the non-spinning instrument. A conservative value   
$\epsilon=20$ $\mu$m is assumed for the offset, and it is checked 
{\it a posteriori} not to have any sizable effect on these results.
\begin{table}[htbp]
\caption{Input parameters for the numerical calculations: geometrical
dimensions of the real bodies. (A mounting error of $\epsilon=20$ $\mu$m has
also been used)}
\vspace{1mm}
\begin{tabular}{llllll}
\colrule
 Body &$m_\lambda$& $R_{\lambda I}$&$R_{\lambda E}$&$L_\lambda$&$R_{\lambda H}$\\
&(kg)&(cm)&(cm)&(cm)&(cm)\\

Arm (a)      &0.3         &3.3                    &3.5                   
&19 &$2L_a+\Delta L$\\
&&&&($\Delta L=-0.1353$)   &\\
Outer cyl. (o)&10          &12.1                   &13.1                  
&$2L_a+L_i+\Delta L$&29.8\\
Inner cyl. (i) &10          &8.0                    &10.9                  
&4.5&21.0\\
\colrule
\end{tabular}
\label{T:1}
\end{table}
\begin{table}[htbp]
\caption{Input  parameters for the numerical calculations:  laminar 
suspensions data. In addition, a conservative 
value ${\mathcal Q}(\nu_s)=510$ has been used  taken  from previous measurements of whirl growth (~\cite{GGG1}, Fig. $7$).}
\vspace{1mm}
\begin{tabular}{llll}
\colrule
Suspension &$l$ &$K$& Anisotropy\\
&(cm)&(dyne/cm)&($\Lambda=K_{Y'}/K_{X'}$)\\

Central      &0.5         &$10^6(l/L_a)^2$                    &2.6\\
Outer cyl. (o)&0.5          &$10^6$                   &1.0 \\
Inner cyl. (i) &0.5          &$10^6$                    &1.0\\
\colrule
\end{tabular}
\label{T:2}
\end{table}

\section{Results: the normal modes}\label{SecResNormal}

We have solved for the eigenvalues $\nu_n$ 
of the matrix $A$ in (\ref{eqMotD}), using 
the system parameters listed in Sec.~\ref{Secsecsysp}. 

\begin{figure}[htbp]
\includegraphics[width=8.5cm]{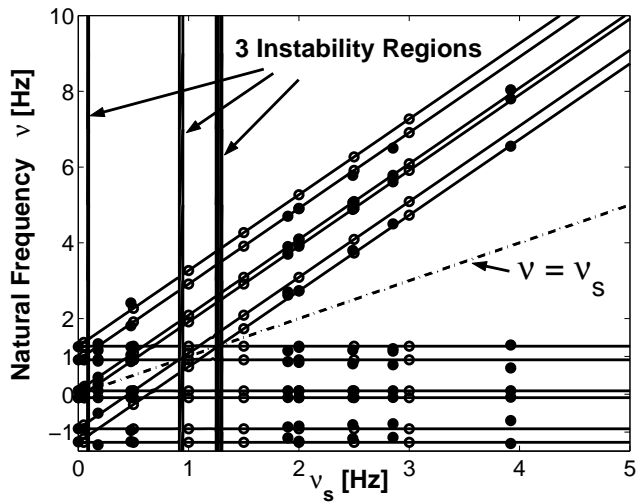}
\caption{Normal modes of the GGG rotor: the frequencies of the 
normal modes are plotted as functions of the spin frequency $\nu_s$. The normal modes as predicted theoretically assuming anisotropic suspensions are shown as $12$ solid lines in the case of zero rotating damping, and as   open circles  in the case of non-zero damping (see text).  The experimental results are plotted as  filled circles, and clearly agree with the theoretical predictions. The  bisecting dot-dashed line $\nu=\nu_s$
separates the supercritical ($\nu_s>\nu$) from the 
subcritical ($\nu_s<\nu$) region. Three vertical 
thick lines are plotted in correspondence of  $3$ instability regions, their thickness referring to the width of the regions 
(see Sec.~\ref{resonances})} \label{fig:normalmodes}
\end{figure}

Fig.~\ref{fig:normalmodes} summarizes our results by plotting the normal modes of the system
 ($\nu_n$ in the 
non-rotating frame) as 
functions of the spin frequency $\nu_s=\omega_s/2\pi$ of the rotor. 
In this figure, theoretical results for $\nu_n(\nu_s)$'s  are displayed by the solid lines in the case of zero rotating damping (i.e., no dissipation in the suspensions), and by the open circles in the case of non-zero  rotating damping (${\mathcal Q}(\nu_s)=510$). At zero damping there are $12$ lines, $6$ horizontal and $6$ inclined; starting from the $3$ natural frequencies (Sec.~\ref{whirl}) of the system, we get $3$x$2$x$2$=$12$ normal mode lines, a factor $2$ being due to anisotropy of the suspensions in the two orthogonal direction of the plane, and the other to the positive and negative sign ($i.e.$ counter-clockwise or clockwise whirl motion). For the non-zero damping case (open circles) 
a conservative low 
value ${\mathcal Q}(\nu_s)=510$ has been assumed (see Tab. ~\ref{T:2}), referring to a comparatively large dissipation. This value has been obtained from previous not so favorable measurements of whirl growth, while much higher ${\mathcal Q}$ values (namely, much smaller dissipations) are expected (see discussion on this issue in~\cite{GGG2}, Sec 3).
In Fig.~\ref{fig:normalmodes}, to be compared with the above theoretical results, we plot, as  filled circles,  the experimental results too,  finding an excellent agreement between theory and experiments. 
Since  Fig.~\ref{fig:normalmodes} contains the  crucial results of this work, it is worth discussing its main features in detail. The main features are: the comparison with the experiment, the role of damping, the behavior at low spin frequencies, the so-called scissors's shape, the splitting of the normal modes and the presence of three instability regions. 

\subsection{Comparison with the experiment}

In the experiment, the rotor is first accelerated to spin at a given 
frequency $\nu_s$.  Then, the natural modes are excited 
by means of capacitance actuators (indicated as $OP$ in Fig.~\ref{Fig:GGGscheme}
in the $X'$ or $Y'$ 
directions at frequencies close to the natural frequencies 
$\nu_n^0\equiv\nu_n(\nu_s=0)$ of the system at zero spin.
 The excitation is performed
for  several 
(typically ten) fundamental cycles $1/\nu_n^0$ by means of voltages applied to $4$ of the $8$ outer plates ($OP$ in Fig.~\ref{Fig:GGGscheme}).
The actuators are then 
switched off, and the bodies' displacements are recorded as functions
of time by means of the read-out described in Sec.~\ref{Secsecread}. 
Standard data analysis is then performed, by fitting the measurement data 
to extract oscillation frequencies and damping of the modes.

Experimental data, resulting from averaging over 
several measurements, are represented as 
filled circles in Fig.~\ref{fig:normalmodes}. The agreement between theory and experiments is excellent, thus 
validating  the model developed 
in Sec.~\ref{SecMod}.

In the experimental spectra as well as in the theory,it is found that the amplitudes of the modes in the subcritical region $\nu_s<\nu$ $-$represented by the inclined lines, with their open and filled circles$-$ are quite small, while non-dispersive modes (the horizontal lines, not varying with the spin frequency) are preferably excited. When the horizontal lines cross the inclined ones, the latter modes can  also be excited.  Since  the excited modes must obviously be avoided in operating the experiment, this information is very useful, in that it is telling us that we should avoid to spin the system at frequencies where  these line crossings occur. Even more so, spin frequencies lying in the instability regions must be avoided  (see Sec. ~\ref{resonances})

\subsection{Role of damping}

We have numerically checked that  dissipation present in the system does not significantly shift the natural-mode frequencies.  This is apparent in Fig.~\ref{fig:normalmodes}, where the results obtained with damping (open circles) stay on the solid lines  obtained in absence of damping. It is worth stressing the this result is especially good because, as discussed above,  we have used a  low value ${\mathcal Q}(\nu_s)=510$, corresponding to comparatively large dissipations As expected, dissipation affects the line-shape of the peaks, making them wider than in absence of damping.

\begin{figure}[htbp]
\hbox{\includegraphics[width=1.7in,height=1.8in]{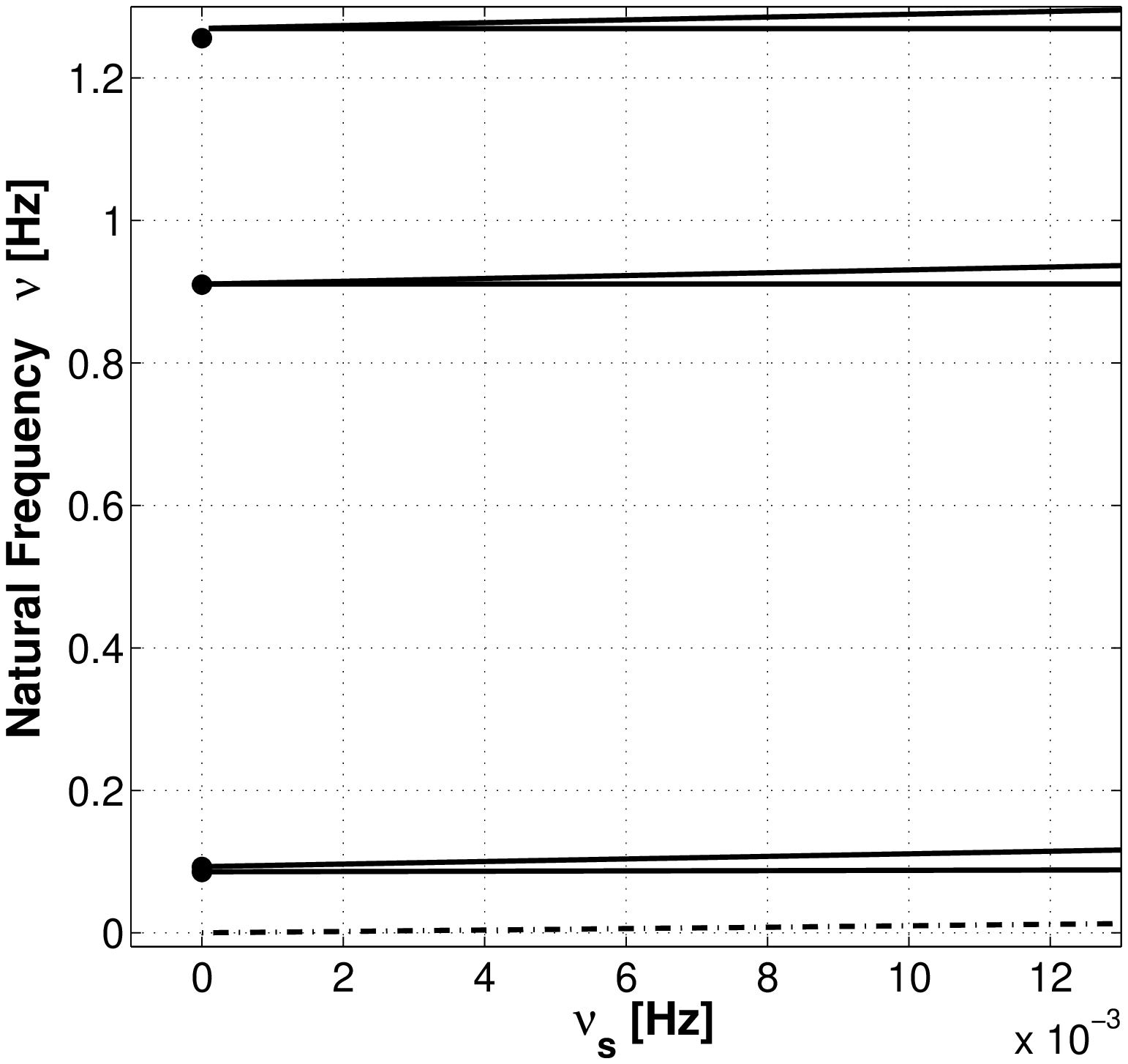}
\includegraphics[width=1.7in,height=1.8in]{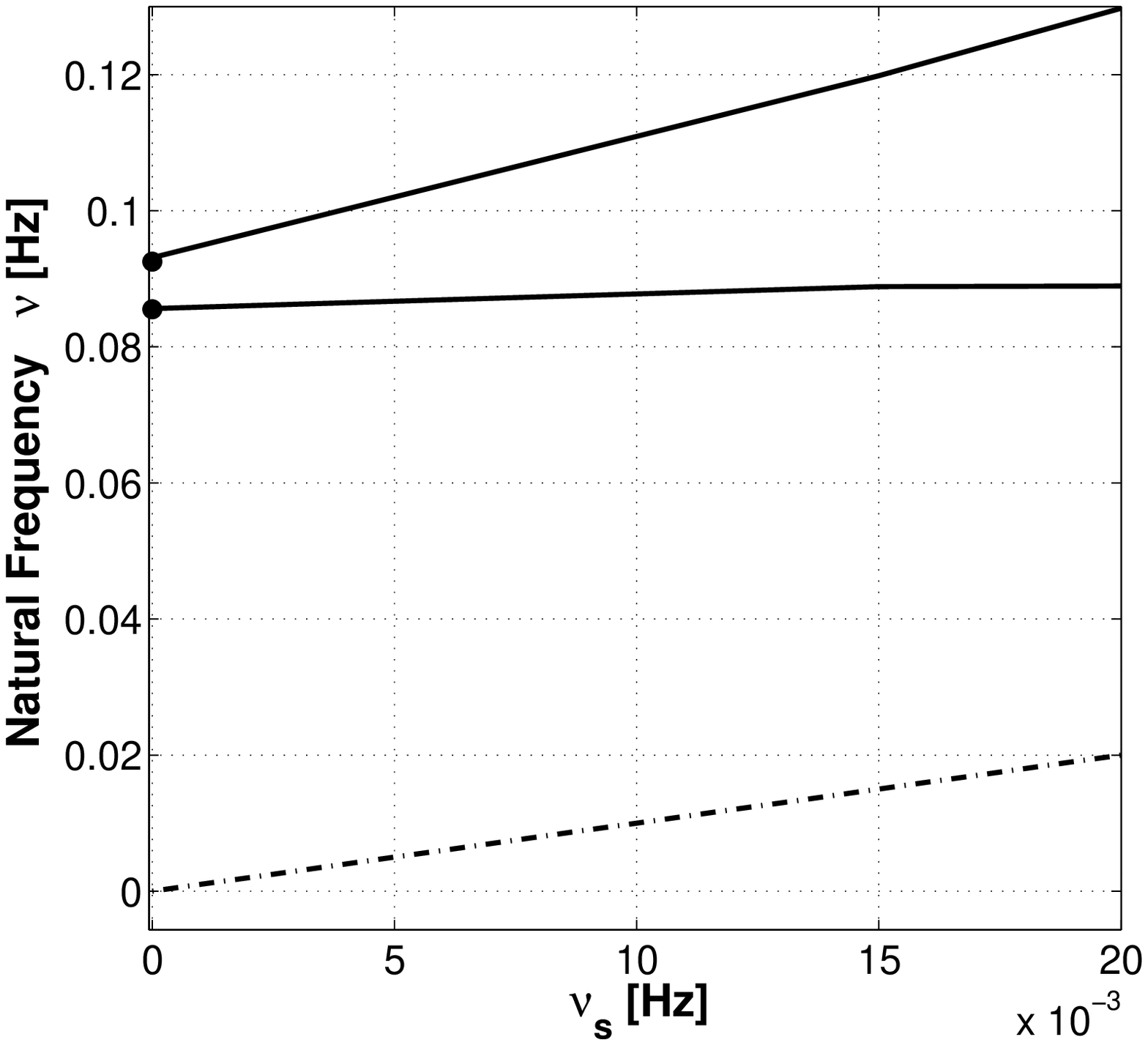}}
\caption{Normal modes of the GGG rotor. On the left hand side we show a zoom from 
Fig.~\ref{fig:normalmodes} in the very low spin frequency region, showing in particular the $3$ natural frequencies of the system in the zero spin case. On the right hand side, we plot a zoom  from 
Fig.~\ref{fig:normalmodes} in the  small frequency region of both axes, showing the splitting into two lines of the low frequency mode because of anisotropy of the suspensions (the dashed line is the  
$\nu_s=\nu$ line as in Fig.~\ref{fig:normalmodes}).} \label{fig:modesplitting}
\end{figure}

\subsection{Low-frequency limit}

At zero spin frequency we have recovered  theoretical and experimental results previously obtained for the non-rotating system. On the left hand side of Fig.~\ref{fig:modesplitting}, a zoom from Fig.~\ref{fig:normalmodes} at very low spin frequencies, we can see that the non-spinning rotor is characterized by three natural frequencies for the instrument with ideally isotropic springs, the three bodies oscillating in a vertical plane. The frequency $\nu^0=0.09\ Hz$ corresponds to the differential mode, where the centers of mass of the two test bodies oscillate in opposition of phase; the frequencies $\nu^0=0.91\ Hz$ and $\nu^0=1.26\ Hz$ correspond to common modes, in which the common center of mass of the two test bodies is displaced from the vertical. During rotation, the number of degrees of freedom increases to six, as discussed in Sec.~\ref{SecsecGenCoo}, leading to the six lines plotted in the same figure.

We may get a flavor of the $\nu_s$ dependence of the modes in the $\nu_s\ll \nu_n$ limit, by evaluating the natural frequencies of only one spinning cylinder with mass $m$, moments of inertia $I_{\xi}$ and $I_\zeta$. The cylinder is  suspended at distance $L$ and with offset $\epsilon$ from a fixed frame by means of a cardanic suspension with isotropic elastic constant $K$ and length $l$. The calculation is performed by following the steps outlined in Sec.~\ref{SecMod} with $n_b=1$ and thus $n=4$. After evaluating the $A$-matrix (Appendix~\ref{appC}, see Eq.~(\ref{eqSLow})) and solving $det(A-sI)=0$, we obtain the two double solutions \begin{equation} \nu_{n1,2,3,4}=\pm\tilde{\nu}_n^0\frac{L}{L'} \label{eqLow} \end{equation} for the four $\nu_n(\nu_s/\nu_n\rightarrow 0)$ in the non-rotating frame. In Eq.~(\ref{eqLow}), $\tilde{\nu}_n^0=(2\pi)^{-1}\sqrt{g/L+K l^2/mL^2}$ is the natural frequency for the non-spinning point-like mass. $L'=\sqrt{L^2+(I_\xi-2I_\zeta)/m}$ (with $I_\zeta<0.5 mL^2+I_\xi$, see Appendix~\ref{appC}), takes into account the extended nature of the body, and the ratio ${L}/{L'}$ modifies $\nu_n$ with respect to $\tilde{\nu}_n^0$.

\subsection{Anisotropy}

If the suspensions  are not isotropic  in the two orthogonal directions, as it is indeed the case for our real cardanic suspensions, each natural frequency is expected to split up. This is clearly shown on the right hand side of Fig.~\ref{fig:modesplitting}, It is worth noting that the splitting is larger for the lowest frequency mode.

\subsection{Scissors's shape}

Fig.~\ref{fig:normalmodes} shows that each natural frequency of the non-spinning system splits up into two branches at $\nu_s>0$, a lower branch remaining approximately constant and an upper branch increasing with $2\nu_s$.

This characteristic scissors's shape can be traced back to general properties of spinning bodies (see also Part II). We again use  the one-cylinder simple case (see Appendix~\ref{appC}) to prove this statement. By following the same procedure which has led to Eq.~(\ref{eqLow}) we obtain (see Eq.~(\ref{eqSHigh})) \begin{eqnarray} \nu_{1,2}&=&\pm\sqrt{\nu_s^2-2\nu_n\nu_s\frac{L}{L'}}\simeq\pm\nu_s\left(1- \frac{\nu_n}{\nu_s}\frac{L}{L'}\right)\label{eqHigha}\\ \nu_{3,4}&=&\pm\sqrt{\nu_s^2+2\nu_n\nu_s\frac{L}{L'}}\simeq\pm\nu_s\left(1+ \frac{\nu_n}{\nu_s}\frac{L}{L'}\right)\label{eqHighb}\; , \end{eqnarray} showing  that in the rotating frame $\nu_n\propto\nu_s$ to zeroth order. After taking the $\nu_s/\nu_n\rightarrow\infty$ limit and tranforming back to the non-rotating frame by means of the substitution $s_n=2\pi i\nu_n\rightarrow 2\pi (i\nu_n+i\nu_s)$,  we finally have \begin{eqnarray} \nu_{1,3}&\simeq&\pm\nu_n\frac{L}{L'}\label{eqHigh1}\\ \nu_{2,4}&\simeq&2\nu_s\label{eqHigh2}\; . \end{eqnarray}
 
\subsection{Mode splitting}

\begin{figure}[htbp] \includegraphics[width=3.2in]{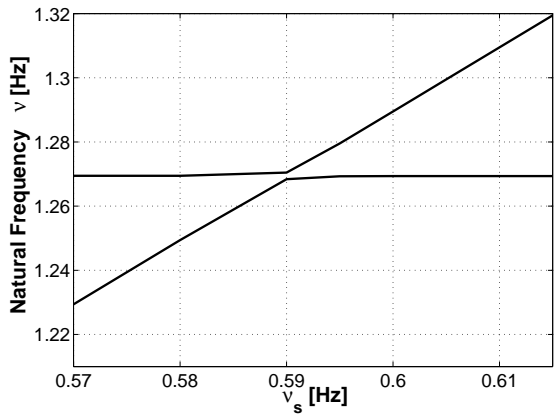} \caption{Normal modes of the GGG rotor. A zoom from Fig.~\ref{fig:normalmodes} showing  one particular case of anticrossing of two modes.} \label{fig:modecrossing} \end{figure}

The two branches may cross at selected frequencies. Crossing and anticrossing of degenerate modes is a very general concept, which applies to a variety of physical systems, from classical to quantum mechanics, from single to many-particle physics. As it is well known~\cite{Weis}, splitting of the modes is expected in correspondence of such crossings. In our numerical results we have found  all the 15 splittings expected for our system (see Fig.~\ref{fig:normalmodes}). Fig.~\ref{fig:modecrossing} shows a particular case of anticrossing of two modes.

\subsection{Instability regions}\label{resonances} 

Dynamical instability may occur whenever the values of the natural frequencies are in proximity of the spin frequency. In such regions the oscillation amplitude grows exponentially.

\begin{figure}[htbp] \includegraphics[width=3.2in]{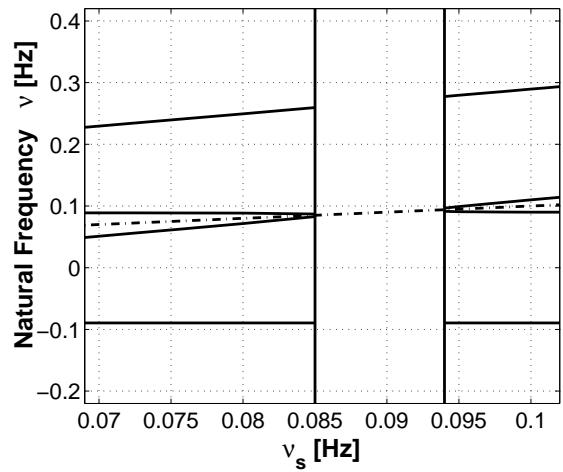} \caption{Normal modes of the GGG rotor. The lowest-frequency instability region is zoomed in from Fig.~\ref{fig:normalmodes}. } \label{fig:instabilityregion} \end{figure}

This is a well-known characteristic of rotating machines; in engineering books it is usually described within the simple model of the so-called Jeffcott rotor~\cite{Genta}. The number of instability regions can be predicted from Fig.~\ref{fig:normalmodes} after drawing the dotted-dashed line $\nu=\nu_s$. We have found indeed three instability regions. Fig.~\ref{fig:instabilityregion} displays in detail the one at the lowest frequency; as shown in Fig.~\ref{fig:normalmodes}, the two at higher frequencies are found to be wider and closer to each other. These theoretical results do explain why in the experiment we can increase the spin frequency and cross the low frequency instability region easily, while it is much more difficult to cross the frequency range  $0.9\rightarrow 1.3\ Hz$. In the past we solved this problem by designing and installing passive dampers to be switched on from remote just before resonance crossing, and then turned off at higher spin frequencies; the least noisy  was a special, no oil damper described in~\cite{GLPhD}, p. 45.  Later on the GGG rotor imperfections have been reduced so that all instability regions can now be crossed, if the crossing is sufficiently fast,  without producing any relevant disturbances even in absence of a passive damper. The physical space previously occupied in the vacuum chamber by the passive damper is now used for the inductive power coupler, indicated as $PC$ in Fig.~\ref{Fig:GGGscheme}, which provides the necessary power to the rotating electronics and has allowed us to avoid noisy sliding contacts.

\section{Concluding remarks}\label{SecCon}

We have demonstrated that the linearized model set up in Sec.~\ref{SecMod} can quantitatively account for the dynamical response of the GGG rotor, an apparatus  designed to test the equivalence principle with fast rotating, weakly coupled,  macroscopic, concentric cylinders (Sec.~\ref{SecExp}). The model developed here  can be expanded to include external disturbances whose effects need to be taken into account in testing the equivalence principle. A qualitative understanding has been provided, by means of helpful analytical solutions of the simplified model under special limits, of relevant features observed in the simulations as well as in the experimental data.

We have  acquired  a detailed knowledge of the instrument's features and the way it works, the main feature being the normal modes of the system (Sec.~\ref{SecResNormal}) in the whole range of spin frequencies, from subcritical to supercritical regime, and as functions of the governing parameters (see  Sec.~\ref{Secsecsysp}).

In particular: we have established the location and characteristics of the instability regions; we have verified quantitatively the effects of dissipation in the system, showing that losses can be dealt with and are not a matter of concern for the experiment; we have established the split up of the normal modes into two scissors's-like branches, distinguishing modes which are preferentially excited (the horizontal lines) from those whose spectral amplitudes are typically small (the inclined lines), thus learning how to avoid the spin frequencies corresponding to  their crossings, in order not to excite the quiet modes too by exchange of energy;   we have investigated  the self-centering characteristic of the GGG rotor when in supercritical rotation regime, gaining insight on how to exploit this very important physical property  for improving the quality of the rotor, hence its sensitivity as differential accelerometer. 

In the following Part II of this work we apply the same model and methods developed here to investigate the common mode rejection behavior of the GGG rotor, a crucial feature of this instrument devoted to detect extremely small differential effects.

\acknowledgments

Thanks are due to INFN for funding the GGG experiment in its lab of San Piero a Grado in Pisa. 

\appendix

\section{The Lagrange function in the rotating reference frame}\label{appA}
In the following, in order to 
simplify the notation we drop the $\lambda$ indices everywhere, and 
restrict our reasoning to only one body.
Let us begin with the expression (\ref{eqTgen}). After using Eq.~(\ref{eqv}) into 
(\ref{eqTgen}), we have 
\begin{equation}
{\cal T}=\frac{1}{2}\int_{\tau_i}\left[\vec{V}^2+\left(
\vec{\Omega}\times\vec{\rho}\right)^2+2\vec{V}\cdot
\left(\vec{\Omega}\times\vec{\rho}\right)\right]dm\; .
\label{eqTapp}
\end{equation}

We conveniently represent the vectors $\vec{\Omega}=\stackrel
{\leftrightarrow}{{\cal M}}
\vec{\Omega}_{\xi\eta\zeta}$ 
and $\vec{\rho}=\stackrel{\leftrightarrow}{{\cal M}}
\vec{\rho}_{\xi\eta\zeta}$
in the 
$\{\xi\eta\zeta\}$ frame by means of the rotation matrix 
\begin{equation}
\stackrel{\leftrightarrow}{{\cal M}}=\left(
\begin{array}{lll}
\sin\phi&\cos\theta\cos\phi&-\sin\theta\cos\phi\\
-\cos\phi&\cos\theta\sin\phi&-\sin\theta\sin\phi\\
0&\sin\theta&\cos\theta\\
\end{array}
\right)\label{eqMRot}
\end{equation}
with respect to the $\{X'Y'Z'\}$ reference system.

Thus, by exploiting the properties of the vectorial product and the definition of 
center-of-mass, namely 
$\int_{\tau_i}\rho_\xi=\int_{\tau_i}\rho_\eta=\int_{\tau_i}
\rho_\zeta=0$, 
we find the following results for the integrals appearing in 
Eq.~(\ref{eqTapp}):
\begin{eqnarray}
\frac{1}{2}\int_{\tau_i}\vec{V}^2dm&=&\frac{1}{2}m\vec{V}^2\nonumber\\
\frac{1}{2}\int_{\tau_i}\left(\vec{\Omega}\times\vec{\rho}\right)^2 dm&=&
\frac{1}{2}\sum_\alpha I_{\alpha\alpha}\Omega_\alpha^2\nonumber\\
\frac{1}{2}\sum_\alpha I_{\alpha\alpha}\Omega_\alpha^2&=&\frac{1}{2}I_\xi\left(
\dot{\phi}^2\sin^2\theta+\dot{\theta}^2\right)\nonumber\\
\int_{\tau_i}\vec{V}\cdot\left(\vec{\Omega}\times\vec{\rho}\right) dm
&=&\left(\vec{V}\times\vec{\Omega}\right)\cdot\int_{\tau_i}
\vec{\rho}dm=0\nonumber\\
\int_{\tau_i}\vec{V}\cdot\left(\vec{\omega}_s\times\vec{R}\right) dm&=&
m\vec{V}\cdot\left(\vec{\omega}_s\times\vec{R}\right)\nonumber\\
\int_{\tau_i}\vec{V}\cdot\left(\vec{\omega}_s\times\vec{\rho}\right) dm
&=&\left(\vec{V}\times\vec{\omega}_s\right)\cdot\int_{\tau_i}\vec{\rho}dm=0
\nonumber\\
\int_{\tau_i}\left(\vec{\Omega}\times\vec{\rho}\right)\cdot
\left(\vec{\omega}_s\times\vec{R}\right) dm&=&
\left(\vec{\omega}_s\times\vec{R}\right)\cdot\int_{\tau_i}
\left(\vec{\Omega}\times\vec{\rho}\right)=0\nonumber\\
\int_{\tau_i}\left(\vec{\Omega}\times\vec{\rho}\right)\cdot
\left(\vec{\omega}_s\times\vec{\rho}\right) 
dm&=&I_\xi\omega_s\dot{\phi}\sin^2\theta+I_\zeta\omega_s^2\cos\theta\nonumber\\
\frac{1}{2}\int_{\tau_i}\left(\vec{\omega}_s\times\vec{R}\right)^2 dm&=&
\frac{1}{2}m
\left(\vec{\omega}_s\times\vec{R}\right)^2\nonumber\\
\frac{1}{2}\int_{\tau_i}\left(\vec{\omega}_s\times\vec{\rho}\right)^2 dm&=&
\frac{1}{2}\left[
I_\xi\sin^2\theta+I_\zeta\cos^2\theta\right]\omega_s^2\nonumber\\
\int_{\tau_i}\left(\vec{\omega}_s\times\vec{R}\right)\cdot
\left(\vec{\omega}_s\times\vec{\rho}\right) 
dm&=&\left(\vec{\omega}_s\times\vec{R}\right)\cdot\int_{\tau_i}
\left(\vec{\omega}_s\times\vec{\rho}\right) 
dm=0\; .\nonumber
\end{eqnarray}
By collecting all these results, one ends up with the final forms 
(19)-(\ref{eqCenden}) for the original ${\cal T}$ 
function~(\ref{eqTgen}). 

\section{The one-cylinder solution}\label{appC}

It is useful to study (along the lines of Sec.~\ref{SecMod}) the 
simplified case of only one 
 spinning cylinder with mass $m$, moments of inertia $I_{\xi}$ and 
$I_\zeta$. This  amounts to setting $n_b=1$ and
thus $n=4$ for the number of generalized coordinates.

The $A$ matrix turns out to be
\begin{equation}
A=\left(
\begin{array}{llll}
0&1&0&0\\
\frac{{\cal L}_{11}}{{\cal L}_{22}}&\frac{{\cal R}_{12}}{{\cal L}_{22}}
&\frac{{\cal R}_{13}}{{\cal L}_{22}}
&\frac{{\cal L}_{14}-{\cal L}_{23}}{{\cal L}_{22}}\\
0&0&0&1\\
\frac{{\cal Q}_{31}}{{\cal L}_{44}}&
\frac{{\cal L}_{23}-{\cal L}_{14}}{{\cal L}_{44}}
&\frac{{\cal L}_{33}}{{\cal L}_{44}}
&\frac{{\cal R}_{34}}{{\cal L}_{44}}\\
\end{array}
\right)\; ,
\label{eqA1}
\end{equation}
where the coefficients of ${\cal L}$ and ${\cal R}$ are defined in terms of 
the system parameters and of the equilibrium positions $\theta_0\ ,\phi_0$ as
\begin{eqnarray}
{\cal L}_{11}&=&m\omega_s^2 L\left(L\cos 2\theta_0+\epsilon\sin\theta_0\right)\nonumber\\
&-&mgL\cos\theta_0-Kl^2\cos 2\theta_0+(I_\xi-I_\zeta)\omega_s^2\cos 2\theta_0
\nonumber\\
&-&I_\zeta\omega_s^2\cos\theta_0\label{defCofA1}\\
{\cal L}_{22}&=&mL^2+I_\xi\nonumber\\
{\cal L}_{33}&=&m\omega_s^2\epsilon L\sin\theta_0\nonumber\\
{\cal L}_{44}&=&(mL^2+I_\xi)\sin^2\theta_0\nonumber\\
{\cal L}_{14}&=&{\cal L}_{41}=I_\zeta\omega_s\sin 2\theta_0-m\omega_s\epsilon L
\cos\theta_0\nonumber\\
&+&m\omega_s^2 L^2\sin 2\theta_0\nonumber\\
{\cal L}_{23}&=&{\cal L}_{32}=-m\omega_s\epsilon L\cos\theta_0\; ,
\end{eqnarray}
and 
\begin{eqnarray}
{\cal R}_{12}&=&-(\Gamma_R+\Gamma_{NR}) L^2\nonumber\\
{\cal R}_{13}&=&-\Gamma_{NR}\omega_s L\epsilon \cos\theta_0\label{defCofA1Q}\\
{\cal R}_{31}&=&-\Gamma_{NR}\omega_s L(L\sin 2\theta_0+\epsilon)\nonumber\\
{\cal R}_{34}&=&-(\Gamma_R+\Gamma_{NR}) L^2\sin^2\theta_0\; .
\end{eqnarray}

In the case of negligible dissipation, the eigenvalue equation $det(A-sI)$ for $s=2\pi i\nu$ reads
\begin{equation}
s^4-s^2(a_{21}+a_{43}+a_{24}a_{42})+a_{21}a_{43}=0\; .
\label{eqSec}
\end{equation}
In the $\omega_s\gg\omega_n$ limit, the equilibrium solutions are
\begin{equation}
\theta_0\simeq \frac{\epsilon L}{{L'}^2}\left[1+\left(\frac{L}{{L'}}\right)^2
\left(\frac{{\tilde{\omega}_n}^{0}}{\omega_s}\right)^2\right], \qquad \phi_0=0\; ,
\end{equation}
where $L'=\sqrt{L^2+(I_\xi-2I_\zeta)/m}$, $I_\zeta<0.5 mL^2+I_\xi$, and
\begin{equation}
\tilde{\omega}_n^0=\sqrt{\frac{g}{L}+\frac{Kl^2}{mL^2}}
\end{equation}
is the natural frequency of the point-like mass. Eq.~(\ref{eqSec}) 
becomes then
\begin{equation}
s^4+2s^2\omega_s^2\left[1+\left(\frac{\omega_n^0}{\omega_s}\right)^2\right]+
\omega_s^4\left[1-2\left(\frac{\omega_n^0}{\omega_s}\right)^2\right]=0\; , 
\label{eqSHigh}
\end{equation}
where we have defined the natural frequency of the cylinders mass as
$\omega_n^0\equiv \left({L}/{{L'}}\right)
{\tilde{\omega}_n^{0}}$. 

In the $\omega_s\ll\omega_n$ limit, the eigenvalues equation becomes
instead
\begin{equation}
s^4+2s^2\left(\frac{L}{{L'}}\right)^2
\tilde{\omega}_n^{02}+
\left(\frac{L}{{L'}}\right)^4\tilde{\omega}_n^{04}=0\; .
\label{eqSLow}
\end{equation}
Eqs.~(\ref{eqSHigh})-(\ref{eqSLow}) are used to derive the 
results (\ref{eqLow})-(\ref{eqHigh2}) in the main text.

\section{The self-centering}\label{SecResSC}

\begin{figure}[htbp] 
\includegraphics[width=8cm]{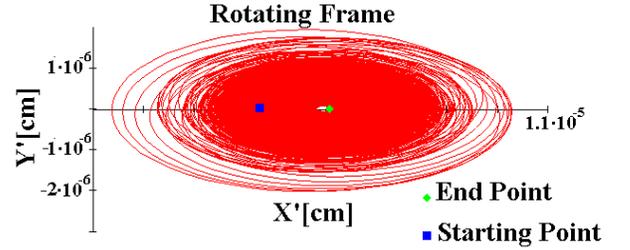} 
\caption{Self-centering of one cylinder  in the presence of non-rotating damping. Simulated $X'(t)-Y'(t)$ plot showing the motion of the center of mass of the cylinder in the rotating reference frame (one cylinder model, $\nu_s=5\; Hz$). The center of mass   spirals inward from the  initial offset value and large initial oscillations to a  final value, much closer to the rotation axis. } 
\label{fig:selfcentering} \end{figure}

This Appendix is devoted to a key feature of the GGG experiment, namely the concept of self-centering of the rotor in supercritical rotation. Let us  analyze the one-cylinder case, by numerically integrating the equations of motion in the presence of  non-rotating damping, to make the rotor asymptotically stable (Sec.~\ref{damping}).  Fig.~\ref{fig:selfcentering} shows the resulting motion of the cylinder in the horizontal plane of the rotating reference frame: its center-of-mass spirals inward towards an equilibrium position much closer to the origin, $i.e.$ to the rotation axis. The equilibrium position always lies in the same direction as the initial offset vector $\vec\epsilon$, which in this simulation was assumed to be in the $X'$ direction. The center of mass of the cylinder  will eventually  perform small-amplitude oscillations around the asymptotic value $\{X'_0=\epsilon-L\sin\theta_0,Y'_0=0\}$.

In the limit of small angles we obtain
\begin{equation}
\theta_0\simeq\pm\frac{\epsilon}{L}\left[\frac{1}{(L'/L)^2-(\omega_n/\omega_s)^2}
\right]\; ,
\label{eqt0}
\end{equation} 
with 
\begin{equation}
\phi_0=0\; (\pi)
\end{equation}
in the case of the lower (upper) sign in Eq.(~\ref{eqt0}), respectively (angles defined as in Fig.~\ref{fig:model}). 
The cylinder's center of mass is eventually located at a
distance
\begin{equation}
\Delta X\simeq\epsilon\pm L\theta_0=
\epsilon-\left[\frac{\epsilon}{(L'/L)^2-(\omega_n/\omega_s)^2}
\right]
\label{eqeccola}
\end{equation}
from the rotation  axis. 

\begin{figure}[htbp] \includegraphics[width=9cm]{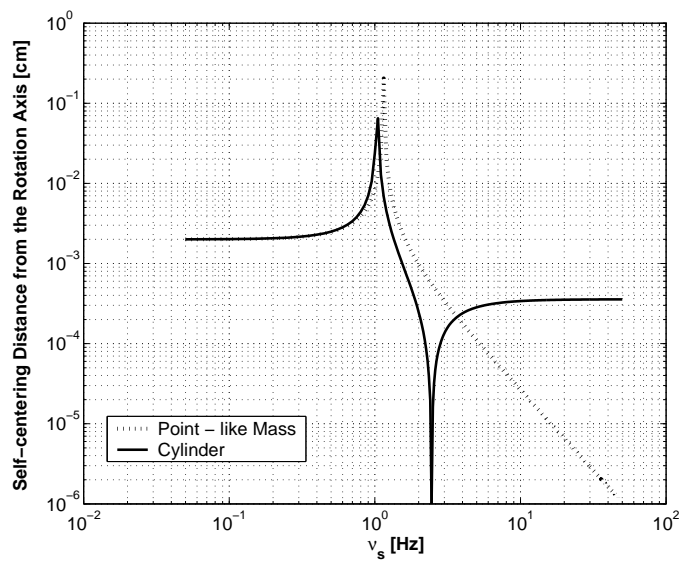} \caption{Self-centering of one cylinder in the presence of non-rotating damping. The distance $\Delta X$ of the center of mass of the cylinder from the rotation axis is plotted as  function of the spin frequency $\nu_s$, in agreement with Eq.~(\ref{eqeccola}). The same distance in  the case of a point mass  is plotted  as  dashed line. See text for comments on their comparison. } \label{fig:selfcentering2} 
\end{figure}

In Fig.~\ref{fig:selfcentering2} we plot, as function of the spin frequency $\nu_s$,  the self-centering distance  $\Delta X$ in the  one-cylinder case  discussed above, and in the   point  masss case. According to the previous Appendix, if $L'\neq L$ we have the cylinder, while if $L=L'$ we have the point mass. The two curves are worth comparing. They have a similar behavior till the resonance peak (in this case, at about $\nu_s\approx 1\; Hz$); the distance from the rotation axis remains constant till, at spin frequencies slightly below the natural one, it starts increasing  showing a typical peak at the resonance. For the cylinder and the point mass the peaks are slightly shifted. The constant value can be obtained 
 from
Eq.~(\ref{eqeccola}) in the limit of small spin frequencies
$\omega_s\ll\omega_n$, finding 
\begin{equation}
\Delta X\simeq \epsilon\; 
\end{equation}

We can also recover the position and relative shift of the resonance peaks in the two cases 
from the values $\nu_s^p$ (the spin frequency at the peak) taken by the 
poles of $\Delta X$ in Eq.~(\ref{eqeccola}), namely 
\begin{equation}
\nu_s^p=\pm\frac{L}{L'}\nu_n\; .
\end{equation} 
Thus, the position of the peaks is dictated by the natural 
frequency $\nu_n$, while the shift is due to the difference 
between $L$ and $L'$. 

At rotation speeds above the resonance, and in highly  supercritical regime $\omega_s\gg\omega_n$, 
the behavior of the cylinder and that of the point mass are remarkably 
different.

For the cylinder,  $\Delta X$ drops to a minimum and then saturates at a  constant value, while  for the point mass it keeps decreasing monotonically.   
The minimum for the cylinder is related to the presence of a zero 
in Eq.~(\ref{eqeccola}), namely
\begin{equation}
\nu_s^z=\pm\frac{L}{\sqrt{L'^2-L^2}}\nu_n\; \; ,
\end{equation} 
which  is valid only if $L'> L$. Instead, for the point mass we have 
the finite value $\Delta X=\epsilon/(1-\omega_s^2/\omega_n^2)$. Note that the
position of the minimum  shifts towards higher spin frequencies as  
 $L'\rightarrow L$, namely, as the finite cylinder case approaches a point mass. In the limit $\omega_s\gg\omega_n$ ($i.e.$ highly supercritical speeds) Eq.~(\ref{eqeccola}) yields
\begin{equation}
\Delta X
\simeq \epsilon\left[1-\left(\frac{L}{L'}\right)^2\right]\; , 
\label{eqSC}
\end{equation}
which explains the saturation to a constant self-centring value in the case of a finite cylinder, whereas a point mass would monotonically approach  perfect centering ($i.e.$, $\Delta X=0$).

In point of fact, it is very interesting to note that $\Delta X$ depends slightly on the point that we are considering along the cylinder's axis. In particular, in the limit $\omega_s\rightarrow\infty$, where $\Delta X$ of the cylinder's center of mass saturates, the point at distance $\tilde{L}=L'^2/L$ from the suspension point along the axis has instead perfect self-centering, namely $\Delta X=0$. This is easily seen from Eq.~(\ref{eqeccola}) after substituting $\epsilon\pm L\theta_0$ with $\epsilon\pm \tilde{L}\theta_0$ and imposing $\Delta X=0$. We plan to exploit this property in order to obtain better self-centering, though it needs further investigation in the actual GGG rotor.

We can consider a plot similar to that of Fig.~\ref{fig:selfcentering2} in the GGG case with two concentric cylinders and a coupling arm, where there are $3$ natural frequencies (one  differential and two common mode).

It happens that the  common mode behavior is similar to that of the one-cylinder case (shown as a solid line in Fig.~\ref{fig:selfcentering2}); namely, for each common mode frequency there is a resonance peak and a minimum peak. Instead,  the differential frequency behavior is similar to that of a point mass (shown as a dashed line in Fig.~\ref{fig:selfcentering2}). This latter fact is  because in the differential mode the coupling arm oscillates and the cylinders' centers of mass move in the horizontal plane with  opposite  phase, while $\theta_{i,o}=0$;  under these conditions, their moment of inertia is irrelevant in determining the dynamics, which therefore is very much alike the case of a point mass. As a result, the $\Delta X$ of the GGG rotor for intermediate values of the spin frequency  is characterized by: one peak at low frequency, in correspondence to the differential mode, and two peaks and two minima in correspondence to the common modes. Instead, in the limit of very low and very high spin frequencies, it has a behavior similar to that displayed in Fig.~\ref{fig:selfcentering2}, depending on the values and directions assumed for the initial offsets of the three bodies.

Thus, in order to obtain the best possible centering of the test cylinders in the GGG rotor, one can either spin at a frequency close to the minima of the common modes, or above both of them, in such a condition that the two cylinders are better centered on their own rotation axes than  both  of them are, together,  in common mode. Self-centering on the rotation axes is very important in order  to reduce rotation noise, because we are dealing with rapidly spinning macroscopic bodies and aiming at measuring extremely small effects. The issue therefore needs careful investigation, and to this end realistic numerical  simulations of the apparatus are an essential tool.

Finally, concerning the use of supercritical rotors for EP testing, it is worth mentioning a frequently asked question:  Would a relative displacement of the test bodies caused by an external force $-$such as that resulting from an EP violation$-$ be reduced by self-centering in supercritical rotation as it happens for the original offset $\epsilon$? The answer is ``No", because the offset vector is fixed in the rotating frame of the system, while an external force gives rise to a displacement of the equilibrium position of the bodies in the non-rotating reference frame. In the presence of such a force, whirl motion will take place around the displaced position of equilibrium. A numerical simulation, showing this important feature is reported and discussed in ~\cite{GG}, PLA paper, p. 176.

\end{document}